\documentclass[12pt]{article}
\usepackage{amsmath}
\usepackage{multirow}
\usepackage{amsfonts}
\usepackage{amssymb,graphics,psfrag}
\usepackage{array,epsfig,multirow,graphicx}
\usepackage{comment}
\usepackage{slashed}
\usepackage{hyperref}
\usepackage{tensor}
\usepackage{booktabs}
\usepackage{colortbl}
\usepackage{hhline}
\usepackage{mathtools}
\usepackage{enumitem}
\usepackage{mathrsfs}  
\usepackage{xfrac}
\usepackage{tikz}
\usepackage{amssymb}
\usepackage[nosort]{cite}
\usetikzlibrary{decorations.markings}
\usetikzlibrary{shapes,arrows}
\usetikzlibrary{decorations.pathreplacing}
\usetikzlibrary{arrows,positioning}
\usepackage{verbatim}
\usepackage{makecell}

\def\hybrid{\topmargin -20pt    \oddsidemargin 0pt
        \headheight 0pt \headsep 0pt 
        \textwidth 6.25in      
        \textheight 9 in      
        \marginparwidth .875in
        \parskip 5pt plus 1pt
          \jot = 1.5ex
  }
\hybrid
\numberwithin{equation}{section}
\numberwithin{table}{section}

\newcommand{\beq}{\begin{equation}}
\newcommand{\eeq}{\end{equation}}
\newcommand{\be}{\begin{equation}}
\newcommand{\ee}{\end{equation}}
\newcommand{\bea}{\begin{eqnarray}}
\newcommand{\eea}{\end{eqnarray}}
\newcommand{\ben}{\begin{eqnarray*}}
\newcommand{\een}{\end{eqnarray*}}               
\newcommand{\ba}{\begin{align}}
\newcommand{\ea}{\end{align}}
\newcommand{\bt}{\begin{tabular}}
\newcommand{\et}{\end{tabular}}
\newcommand{\bc}{\begin{center}}
\newcommand{\ec}{\end{center}}
\newcommand{\ax}{\alpha}
\newcommand{\bx}{\beta}

\newcommand{\dx}{\delta}

\newcommand{\cO}{\mathcal{O}}

\newcommand{\cR}{\mathcal{R}}

\newcommand{\nn}{\nonumber}

\newcommand{\cref}{{\bf [check ref]}}

\newcommand{\tr}{\mathrm{tr}}

\definecolor{mppgreen}{RGB}{17,102,86}
\definecolor{mppgray}{RGB}{221,222,214}

\def\tbz{{\scriptscriptstyle{(0)}}}

\newcommand{\sst}[1]{\scriptstyle{#1}}

\def\upd{\mathrm{d}}

\def\om{\omega}

\def\fr{\frac}

\def\Z0{Z^\tbz}

\def\t{\text}
\def\ads{{\text{AdS}{}_3}}

\def\w{\wedge}
\def\tfr{\tfrac}
\def\Om{\Omega}
\def\tr{\text{tr}\,}
\def\cR{\mathcal{R}}

\def\l{\ell}

\def\w{\wedge}
\def\cR{\mathcal{R}}

\def\lab{\label}
\def\tbsix{\scriptscriptstyle{(6)}}
\def\sst{\scriptscriptstyle}

\def\blfootnote{\xdef\@thefnmark{}\@footnotetext}
\long\def\symbolfootnote[#1]#2{\begingroup%
\def\thefootnote{\fnsymbol{footnote}}\footnote[#1]{#2}\endgroup}

\begin{document}

\baselineskip=15pt

\begin{titlepage}
\begin{flushright}
\parbox[t]{1.8in}{\begin{flushright} ~ \\
~ \end{flushright}}
\end{flushright}

\begin{center}

\vspace*{ 1.2cm}

{\LARGE Four-dimensional black hole entropy from F-theory}

\vskip 1.2cm

\renewcommand{\thefootnote}{}
\begin{center}
 {Thomas W.~Grimm, Huibert het Lam, Kilian Mayer and Stefan Vandoren \footnote{t.w.grimm@uu.nl ~ \ ~ h.hetlam@uu.nl~ \ ~ k.mayer@uu.nl~ \ ~  s.j.g.vandoren@uu.nl}}
\end{center}
\vskip .2cm
\renewcommand{\thefootnote}{\arabic{footnote}}

{
Institute for Theoretical Physics and \\
Center for Extreme Matter and Emergent Phenomena,\\
Utrecht University, Princetonplein 5, 3584 CE Utrecht, The Netherlands\vspace{0.5cm}
}

\vspace*{.2cm}

\end{center}

 \renewcommand{\thefootnote}{\arabic{footnote}}
 
\begin{center} {\bf Abstract } \end{center}

We study the central charges and levels of a two-dimensional $N=(0,4)$ superconformal field theory describing four-dimensional BPS black holes in F-theory. These arise from D3-branes wrapping a curve in the base of an elliptically fibered Calabi--Yau threefold times a circle, and probe a transverse Taub-NUT space. The near horizon geometry of these D3-branes is AdS$_3 \times \text{S}^3/\mathbb{Z}_m$, where $m$ is the NUT charge. Starting from a six-dimensional supergravity effective action we compute three-dimensional Chern-Simons terms to deduce the central charges and levels. We find that it is crucial to integrate out an infinite tower of massive Kaluza-Klein states on S$^3/\mathbb{Z}_m$ to match the expected microscopic results. The induced corrections turn out to contribute at leading order to the central charges and levels, which in turn determine the black hole entropy.

\end{titlepage}

\tableofcontents

\newpage



\section{Introduction}

Ever since the breakthrough in \cite{Strominger:1996sh}, which provided the first microscopic derivation of the Bekenstein-Hawking entropy of BPS black holes, a huge number of works have been devoted to match microscopic and macroscopic entropy calculations. Along with the original reference \cite{Strominger:1996sh} which studied five-dimensional non-spinning black holes in compactifications of type II strings on T$^5$ and K$3 \times \mathrm{S}^1$, the most prominent examples are five-dimensional non-spinning black holes in compactifications of M-theory on a Calabi-Yau threefold CY$_3$ \cite{Vafa:1997gr}, and four-dimensional black holes in M-theory on CY$_3 \times$ S$^{1}$ \cite{Maldacena:1997de}. Another active area of research in the past decade is F-theory \cite{Vafa:1996xn}, a geometrized formulation of type IIB string theory with varying string coupling. F-theory turned out to be particularly powerful for studying models of particle physics and geometric engineering of gauge theories in various dimensions, for nice reviews see e.g.~\cite{Weigand:2018rez,Taylor:2011wt}. The focus in these constructions was on extracting information such as gauge groups, charged matter spectra and Yukawa couplings from the F-theory compactification geometry. However, interesting gravitational features like black holes and black branes, remained comparably less explored.

In the early days of F-theory the microscopic entropy of a D3-brane in an asymptotic geometry $\mathbb{R}^{1,4} \times \mathrm{S}^1 \times \mathrm{CY}_3$ was computed exploiting its dual M-theory formulation. The D3-brane is wrapped on $\t{S}^1\times C$, where $C$ is a curve in the base $B$ of an elliptically fibered Calabi-Yau threefold CY$_3$ and corresponds to a non-spinning black hole in five dimensions. The microscopic entropy was then successfully matched to its macroscopic counterpart \cite{Vafa:1997gr}. Some years later, this setup was generalized to an asymptotic geometry $\mathbb{R} \times \mathrm{S}^1 \times \t{TN}_m  \times \mathrm{CY}_3$ \cite{Bena:2006qm} which corresponds to macroscopic four-dimensional black holes. The microscopic analysis was carried out by mapping the F-theory setup to the MSW setting \cite{Maldacena:1997de}. Studying gravitational aspects in F-theory attracted renewed interest recently using diverse approaches. In \cite{Haghighat:2015ega} the authors extended the study of \cite{Vafa:1997gr} to the case of five-dimensional spinning black holes that previously only had been studied in compactifications of type II strings on T$^5$ and K$3 \times \mathrm{S}^1$ \cite{Breckenridge:1996is}. On the CFT side the main difference between \cite{Vafa:1997gr} and \cite{Haghighat:2015ega} is the identification of an $\mathfrak{su}(2)_L$ current algebra in spite of the absence of supersymmetry in the left-moving sector. Most recently, supersymmetric AdS$_3$ solutions of type IIB supergravity with varying axio-dilaton profile and five-form flux dual to $N=(0, n)$, $n=2,4$ SCFTs were analyzed in \cite{Couzens:2017way,Couzens:2017nnr}. The AdS$_3$ solutions dual to $N=(0,4)$ SCFTs can be interpreted as near horizon geometries of six-dimensional strings from wrapped D3-branes, as described above.

In this paper we derive characteristic data of the SCFT corresponding to the D3-brane wrapped 
inside the $\mathbb{R} \times \mathrm{S}^1 \times \t{TN}_m  \times \mathrm{CY}_3$ geometry \cite{Bena:2006qm} from macroscopic considerations. The two-dimensional SCFT has $N=(0,4)$ supersymmetry and left- and right-moving central 
charges $c_{\,{L, R}}$, as well as left- and right-moving current algebras ${U}(1)_{ L}/ \mathbb{{Z}}_m \times {SU}(2)_{ R}$ 
with levels $k_{{L, R}}$.  The consideration of this setting is strongly motivated by the 4D/5D black hole correspondence 
\cite{Gaiotto:2005gf,Behrndt:2005he}. Making the radius of the circle S$^1$ very small, we perform 
a T-duality along $\mathrm{S}^1$ to obtain a type IIA setting that then lifts to an M-theory 
background $\mathbb{R}  \times \t{TN}_m  \times \mathrm{CY}_3$. Under this duality the wrapped D3-brane turns into an M2-brane 
wrapping $C$. Momentum $n$ along S$^1$ corresponds to bound states of M2-branes wrapping a curve in the class 
$C+n \mathbb{E}_\tau$, where $\mathbb{E}_\tau$ is the elliptic fiber of the CY$_3$. After compactification on 
CY$_{3}$ one obtains a five-dimensional black hole with a transverse Taub-NUT spacetime. This five-dimensional black 
hole has an eigenvalue
$J_L$ which corresponds to the $U(1)_L/\mathbb{Z}_{m}$ symmetry along
the NUT-circle. Compactifying the M-theory setting further along the circle fiber of the Taub-NUT space results in a type IIA compactification on CY$_3$. The M2-brane configuration get mapped to a D6-D2-D0 system on the same Calabi-Yau threefold. The D6-brane has
multiplicity $m$ and one has $2 J_L$ units of D0-brane charge. This is the four-dimensional side of the correspondence in \cite{Gaiotto:2005gf,Behrndt:2005he}. 

Instead of using the circle S$^1$ wrapped by the D3-brane to go to five dimensions, we can also reduce along the Taub-NUT circle, which we denote by $\tilde{\text S}^1$. Performing a T-duality along $\tilde{\text S}^1$, and lifting to M-theory, the D3-brane wrapping the curve $C$ turns into an M5-brane wrapping $\hat C=\pi^{-1}(C) \subset$ CY$_3$, where $\pi: \text{CY}_3 \to B$ is the projection to the base. Following the same duality the Taub-NUT space gives rise to $m$ M5-branes wrapping the base $B$ of the elliptic fibration. The two groups of M5-branes can be combined into a single M5-brane wrapping the curve $\hat C+m B$ in the CY$_3$ if the corresponding class is very ample.
In summary, the two main dualities just introduced can be depicted schematically as:
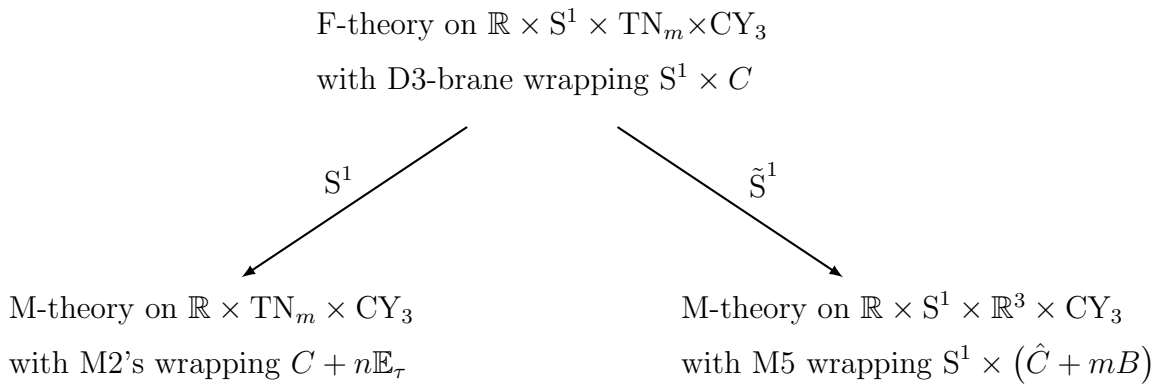
\begin{center}
\begin{tikzpicture}[>=latex]
 \node[align=left ] at (0.8,0) {F-theory on $\mathbb{R} \times \mathrm{S}^1 \times \text{TN}_m \times $CY$_3$\\[0.2cm]with  D3-brane wrapping S$^1 \times C$};
 \draw[->, thick] (-0.2,-1)--(-3.2,-3);
 \draw[->, thick] (1.8,-1)--(4.8,-3);
 \node at (-1.9,-1.7) {${\text{S}^1}$};
 \node at (3.75,-1.7) {${\tilde{\text{S}}}^1$};
 \node[align=left] at (-3.6,-3.8) {M-theory on $\mathbb{R} \times \text{TN}_m \times \text{CY}_3$\\[0.2cm] with M2's  wrapping $C+ n \mathbb{E}_\tau$};
 \node[align=left] at (5.8,-3.8) {M-theory on $\mathbb{R} \times \text{S}^1 \times \mathbb{R}^3 \times \text{CY}_3$\\[0.2cm]
with M5 wrapping $\text{S}^1 \times \big(\hat C + m B \big)$};
 \end{tikzpicture}
 \end{center}
 
If we take both the NUT circle and the circle wrapped by the D3-brane to be small, we obtain an effectively four-dimensional black hole. We therefore obtain an F-theory description of a four-dimensional black hole. The central charges and levels we determine in this paper then give the black hole entropy via the Cardy formula.
 
We use the six-dimensional effective $N=(1,0)$ supergravity action of F-theory compactified on an elliptically fibered Calabi-Yau threefold derived in \cite{Ferrara:1996wv, Bonetti:2011mw} to determine the contributions from classical six-dimensional supergravity to the central charges and levels using techniques of \cite{Kraus:2005vz,Kraus:2005zm,Hansen:2006wu}. Concretely, we dimensionally reduce the six-dimensional effective action to three dimensions and read off the sought-after quantities from the coefficients of Chern-Simons terms. It turns out that in order to fully reproduce the microscopic quantities one also has to include one-loop Chern-Simons terms in three dimensions. These one-loop induced terms arise from integrating out massive Kaluza-Klein modes. This interplay between classical and quantum contributions to complete M/F-theory duality in this case is in fact not unexpected. Including one-loop corrections was already crucial for the matching of the five-dimensional M-theory effective action on CY$_3$ and its dual six-dimensional F-theory action \cite{Bonetti:2011mw,Bonetti:2013ela}. We furthermore utilize the procedure to do the dimensional reduction `at asymptotic infinity' put forward in \cite{Dabholkar:2010rm} instead of doing it in the near horizon geometry, as may be the more intuitive approach in view of the standard AdS/CFT dictionary. As a last ingredient for the comparison with the microscopic charges we also take into account a shift in the charges stemming from a non-vanishing higher-derivative term on Taub-NUT space in the six-dimensional effective action. The latter two points demonstrate that the full geometry outside the horizon is important for the matching with microscopics.

In section \ref{sec: microscopics} we start with a more extensive description of the setting we are working in. In this section we will also state the microscopic quantities that we want to reproduce from supergravity. We then proceed by calculating the classical and quantum contributions to the central charges and levels in sections \ref{sec:classical} and \ref{sec:quantum} respectively. Subsequently we summarize and comment on the 4D/5D correspondence in section \ref{sec: comp with earlier work}. Finally, we discuss our results in section \ref{sec: conclusion}.

\section{Microscopics \lab{sec: microscopics}}
As already stated in the introduction, we consider an F-theory background $\mathbb{R} \times \mathrm{S}^1 \times \t{TN}_m  \times \mathrm{CY}_3$, where  we have a D3-brane wrapping S$^1 \times C$ with $C \subset B$ a curve in the elliptically fibered Calabi-Yau threefold $\pi: \mathrm{CY}_3 \rightarrow B$. For simplicity we only consider threefolds with mild fiber degenerations which render the total elliptic fibration smooth. Using a basis  $\om_\ax$ of $H^{1,1}(B)$ we can expand the Poincar\'{e} dual of the curve and the first Chern class of the base such that we have $C=q^\ax \om_\ax$ and $c_1(B)=c^\ax \om_\ax$. The intersection numbers on the base are given by
\be
 \eta_{\ax \bx} \equiv \int_B \om_\ax \w \om_\bx \, .
\ee
Furthermore, here and in the following we make use of the notation
\begin{align}
C \cdot C&=\int_{{B}}C \w C \equiv C^2\, ,\nn \\
c_1(B) \cdot C&=\int_{B}c_1(B) \w C\, ,\\
c_1(B) \cdot c_1(B)&=\int_{B}c_1(B) \w c_1(B)\equiv c_1(B)^2\, .\nn
\end{align}

Microscopically the central charges corresponding to this setting were derived by considering the dual system in M-theory \cite{Bena:2006qm}. As already described in the introduction one can start from type IIB, T-dualize along the NUT-circle and then lift the system to M-theory.
Performing a T-duality along the NUT circle $\tilde{\text{S}}^1$ we end up with a D4-brane wrapping S$^1 \times \tilde{\text{S}}^1 \times C$ and $m$ NS5-branes wrapping $B$. These type IIA objects lift in M-theory to an M5-brane wrapping $\hat C$ and $m$ M5-branes wrapping $B$. If the class of the curve $\hat C+m B$ is very ample the two M5-brane groups can be combined into a single M5-brane wrapping $\hat C+ m B$. We therefore assume that $q^\ax>0$ and $q^\ax \gg m c^\ax, ~~\forall \ax=1, \dots, h^{1,1}(B)$.

This system falls in the class of settings studied by MSW \cite{Maldacena:1997de} such that the central charges and right level are given by\footnote{These results follow straightforwardly from the data given in \cite{Bena:2006qm} using identities valid for elliptically fibered Calabi-Yau threefolds.}
\begin{align}
c_{L} & =  3mC^{2}-3m^{2}c_{1}(B)\cdot C+m^{3}c_{1}(B)^{2}+12c_{1}(B)\cdot C+12m-2mc_{1}(B)^{2}, \label{central charges} \\
c_{R} & =  6k_{R}=3mC^{2}-3m^{2}c_{1}(B)\cdot C+m^{3}c_{1}(B)^{2}+6c_{1}(B)\cdot C+6m-mc_{1}(B)^{2}, \nn 
\end{align}
where the relation between the right central charge and level follows from supersymmetry in the right-moving sector of the SCFT. 
\paragraph{Left level.} \label{sec:left level}
Although not explicitly calculated, we can extract the left level $k_{L}$ from the data provided in  \cite{Bena:2006qm}. The formula for the entropy of the black string in \cite{Bena:2006qm} reads
\begin{equation}
S=2\pi\sqrt{\frac{c_{L}}{6}\hat{m}}\, ,
\end{equation}
where 
\begin{equation}
\hat{m}=n+\frac{1}{12}\left(D^{00}\tilde{Q}_{0}\tilde{Q}_{0}+2D^{0\alpha}\tilde{Q}_{0}\tilde{Q}_{\alpha}+D^{\alpha\beta}\tilde{Q}_{\alpha}\tilde{Q_{\beta}}\right)\label{eq:momentum}
\end{equation}
and the matrix $D$ is given by 
\begin{align}
D_{00} & =  \frac{1}{6}c_1(B)\cdot C\, , \qquad 
D_{0\alpha}  =  \frac{1}{6}q_{\alpha}\, ,\qquad
D_{\alpha\beta}  =  \frac{1}{6}m\eta_{\alpha\beta}\,.
\end{align}
The elements $D^{AB}=D^{-1}_{AB}$ denote components of the inverse matrix with respect to the full matrix $D_{AB}$, in particular, it is not the inverse of a sub-matrix of $D$. The charge $\tilde{Q}_{0}$ contains a term $2J_{L}/m$ \cite{Bena:2006qm}. The $U(1)$ current $J_L$ belongs to an $SU(2)$ current algebra and the entropy of such a CFT is given by
\begin{equation}
S=2\pi\sqrt{\frac{c_{L}}{6}\Big(n-\frac{J_{L}^{2}}{k_{L}}\,\Big)}\, , \label{cardy formula}
\end{equation}
and we can read off $k_{L}$. The level should not depend on the momentum, so we can take the limit $2J_{L}/m\rightarrow\infty$
in  (\ref{eq:momentum}) and compare the resulting expression with
the spectral flow invariant $n-\frac{J_{L}^{2}}{k_{L}}.$ In particular
we have that
\begin{equation}
\frac{1}{12}D^{00}\left(\frac{2J_{L}}{m}\right)^{2}=-\frac{J_{L}^{2}}{k_{L}}\,.
\end{equation}
Calculating the inverse of $D$ yields 
\begin{equation}
D^{00}=\frac{6m}{m c_1(B)\cdot C-C^2}\, ,
\end{equation}
such that we find
\begin{equation}
k_{L}=\frac{1}{2}mC^2-\frac{1}{2}m^{2}c_1(B)\cdot C \label{left level}
\end{equation}
for the left level. 

\paragraph*{Goal of the paper.} \label{sec:goal}

It is the main objective of this paper to reproduce the central charges and levels, given in (\ref{central charges}) and (\ref{left level}), from six-dimensional (1,0) supergravity up to $\mathcal{O}(1)$ terms.

As described in the introduction, our setting is motivated by four-dimensional black holes. Using $c_L$ and $k_L$, one can compute the entropy of this black hole in the Cardy limit via the formula (\ref{cardy formula}). In the same limit the Wald entropy \cite{Wald:1993nt} is equal to this formula with $c_L$ and $k_L$ derived from the supergravity action \cite{Saida:1999ec,Kraus:2005vz}.

Parts of the central charges and levels have been computed in \cite{Couzens:2017way} from type IIB supergravity. The authors studied AdS${}_3$ solutions of type IIB supergravity with varying axio-dilaton using the spinorial geometry approach. They studied the constraints on the compact geometry arising from preserving $N=(0,4)$ supersymmetry in the dual two-dimensional SCFT while preserving all AdS${}_3$ isometries. The class of ten-dimensional solutions takes the form AdS$_3 \times \t{S}^3/\mathbb{Z}_m \times (\mathbb{E}_\tau \hookrightarrow \t{CY}_3 \overset{\pi}{\rightarrow} B)$, with non-trivial five-form flux and axio-dilaton profile, where $B$ is the K\"ahler base of an elliptically fibered Calabi-Yau threefold. The solution can be interpreted as the near horizon limit of $N$ D3-branes wrapping a curve $C$ in the K\"ahler base in the presence of D7-branes and a Taub-NUT space with NUT-charge $m$ in the four non-compact directions transverse to the D3-branes. The dual $N=(0,4)$ SCFT has again a $U(1)_L/\mathbb{Z}_m \times SU(2)_R$ current algebra with levels $k_{L, R}$. These levels and the central charges of the CFT were computed in the large $N$ limit and were for general $m$ found to be
\begin{align}
c^{\t{IIB}}_R&=6 k^{\t{IIB}}_R=3N^2 \, m\, C^2\, ,\nn\\
c^{\t{IIB}}_L&=3N^2  m\, C^2\, , \lab{IIBresults}\\
k_L^{\t{IIB}}&= \,  \t{unknown}\, .\nn
\end{align}
The subleading correction $c^{\t{IIB}}_L-c^{\t{IIB}}_R$ at $\cO(N)$ was also found for general $m$, it is however expected from the dual M-theory result (\ref{central charges}) that there exist additional $\cO(N)$ contributions to the central charges and levels. The full answer for $c_{L, R}^{\t{IIB}}$ and $k^{\t{IIB}}_R$ including $\cO(N)$ contributions was given for the distinguished case $m=1$, where the near horizon geometry is AdS$_3 \times \t{S}^3 \times B$ corresponding to an unbroken $SU(2)_L \times {SU}(2)_R$ current algebra in the CFT.

\section{Macroscopics in F-theory from 6D: classical contributions} \lab{sec:classical}
In this section we use six-dimensional (1,0) supergravity \cite{Romans:1986er,Ferrara:1997gh,Nishino:1997ff,Sagnotti:1992qw} to compute parts of the microscopic central charges and levels (\ref{central charges}) and (\ref{left level}). An F-theory compactification on a smooth elliptically fibered Calabi-Yau threefold results in a gravity multiplet, $n_T=h^{1,1}(B)-1$ tensor multiplets and $n_H=h^{2,1}(\mathrm{CY}_3)+1$ hypermultiplets, but no vector multiplets \cite{Vafa:1996xn,Morrison:1996na,Morrison:1996pp}. Recall that we restricted ourselves to smooth threefolds for simplicity, e.g~to avoid charged matter. We reproduce part of the central charges and levels utilizing the approach used in \cite{Kraus:2005vz,Kraus:2005zm,Hansen:2006wu,Dabholkar:2010rm}, which in principle means that one has to reduce the six-dimensional action on the spherical part of the near horizon geometry AdS$_{3} \times \t{S}^{3}/\mathbb{Z}_{m}$ of the black string solution. Dimensionally reducing the six-dimensional action on S$^3/\mathbb{Z}_m$ one can infer the levels and central charges of the dual CFT from coefficients of Chern-Simons terms in three dimensions using the AdS/CFT dictionary, see e.g.~\cite{Kraus:2006wn}. In fact we find, based on \cite{Dabholkar:2010rm},  that one has to do this dimensional reduction at spatial infinity of the solution to get the correct result for central charges and levels and to take into account the effect of the Taub-NUT space transverse to the string.

We first provide a few details about the six-dimensional $N=(1,0)$ supergravity theory arising from F-theory compactified on a Calabi-Yau threefold, which shall be the starting point for our investigation. In the sequel we perform the dimensional reduction of the supergravity action to three dimensions, pointing out the difference between the reduction in the near horizon geometry and the reduction at asymptotic infinity. In both cases one finds a mismatch with the microscopic prediction. The mismatch in the reduction at asymptotic infinity can be cured using one loop induced Chern-Simons terms in three dimensions. This will be the subject of section \ref{sec:quantum}, which is one of the main results of this paper.

\subsection{Six-dimensional $N=(1, 0)$ supergravity from F-theory on CY${}_3$}

We consider the six-dimensional effective action arising from compactifying F-theory on an elliptically fibered Calabi-Yau threefold CY$_3$. The characteristic data of the underlying $N =(1,0)$ supergravity in six dimensions was determined in \cite{Bonetti:2011mw} by matching a generic circle reduced six-dimensional $N=(1,0)$ theory with the geometric data arising from compactifying M-theory on a smooth Calabi-Yau threefold. The massless spectrum assembles itself in representations of the little group in six dimensions $SO(4) \simeq SU(2)_1 \times SU(2)_2$, which are labelled by the spins $(j_1, j_2)$. We will focus on a six-dimensional theory with field content 
\begin{itemize}
\item one gravity multiplet: $(1,1)\oplus 2\,(1,\tfr{1}{2}) \oplus (1,0)$ i.e.~one graviton, one left-handed gravitino and one self-dual rank two tensor
\item $n_T$ tensor multiplets: $(0,1) \oplus 2\,(0,\tfr{1}{2}) \oplus (0, 0)$ i.e.~one anti-self-dual rank two tensor, one right-handed tensorino and one real scalar
\item$n_H$ uncharged hypermultiplets: $2\, (0,\tfr{1}{2}) \oplus 4\,(0,0)$ i.e.~one right-handed hyperino and two complex scalars.
\end{itemize}
We will furthermore assume throughout the paper that the six-dimensional spectrum satisfies the anomaly constraint
\be
n_H=273-29n_T\, .
\ee
\paragraph{Tensor multiplets.}

\noindent The rank two tensors in the gravity- and tensor multiplets are collectively denoted by $ \hat B^\ax$ with $\ax=1, \dots, n_T+1$. The scalars in the tensor multiplets parametrize the manifold
\be
\mathcal{M}_{\rm tensor}=\fr{SO(1, n_T)}{SO(n_T)}\, .
\ee
The scalar sector of the tensor multiplets is usually described by $n_T+1$ scalar fields ${\hat \jmath}^{ \, \ax}$ subject to the constraint
\be
\Om_{\ax \bx} \, {\hat \jmath}^{ \,\ax} \, {\hat \jmath}^{ \,\bx}=1\, ,\lab{constr1}
\ee
where $(\Om_{\ax \bx})$ is the $SO(1, n_T)$ invariant constant metric with mostly minus signature. In the six-dimensional F-theory models we consider here this matrix $\Om_{\ax \bx}$ is identified with the intersection numbers on the base, i.e.
\be
\Om_{\ax \bx} = \eta_{\ax \bx}\, .
\ee
The constraint $\eqref{constr1}$ is the six-dimensional analogue of the cubic constraint in very special geometry, which governs the vector multiplet sector in five-dimensional $N=2$ supergravity. One furthermore introduces the non-constant, positive-definite metric
\be
 {g}_{\ax \bx}=2 \, {\hat \jmath}_{\, \ax}\,  \hat \jmath_{ \, \bx}-\eta_{\ax \bx}\, , \qquad  \hat \jmath_{\, \ax} \equiv \eta_{\ax \bx} \,  \hat \jmath^{ \, \bx}\, .\lab{gmet}
\ee
The gauge-invariant field-strength $\hat G^\ax$ is defined by
\be
\hat  G^\ax= \upd  \hat B^\ax+\tfr{1}{8} \, c^\ax \, \hat \om^{\rm CS}_{\rm grav} \, , \qquad  \hat \om^{\rm CS}_{\rm grav} =\tr \big(  \hat \om \wedge \upd \hat \om+\tfr{2}{3}\, \hat \om \wedge  \hat \om \wedge  \hat \om \big)\, ,
\ee
where $\hat \om$ is the six-dimensional spin connection. 

\paragraph{Hypermultiplets.} Every hypermultiplet contains four real scalars, such that we denote the scalars collectively by $\hat q^U$ ($U=1, \dots, 4 n_H$). The hypermultiplets have a geometric interpretation as  coordinates on a quaternionic manifold, whose metric is denoted by $h_{UV}$. Since we do not include vector multiplets in our setting and therefore the hypermultiplets are neutral, we will not need any further information about the hypermultiplets.

\paragraph{Standard form of 6D (1,0) supergravity}

We choose conventions $\kappa_6^2=(2\pi)^3$  such that the bosonic part of the standard $N=(1,0)$ supergravity theory takes the form \cite{Ferrara:1996wv,Bonetti:2011mw}
\begin{align}
S^{\tbsix}=\fr{1}{(2\pi)^3}\int_{M_6} \Big[ \tfr{1}{2} \,  \hat R \,  \hat \ast \, 1&-\tfr{1}{4} \,  g_{\ax \bx}\,  \hat G^\ax \wedge  \hat  \ast \, \hat G^\bx-\tfr{1}{2} \,  g_{\ax \bx}\, \upd  \hat \jmath^{\, \ax} \wedge \hat \ast \, \upd \hat \jmath^{\, \bx}- h_{UV} \, \upd \hat q^U \wedge \hat \ast \, \upd \hat q^V\nn\\[0.2cm]
&-\tfr{1}{8}\eta_{\ax \bx} \,c^\ax \hat B^\bx \wedge  \tr \hat \cR \w \hat \cR \Big]\, .\lab{6dgeneral}
\end{align}
The last term in \eqref{6dgeneral} is a Green-Schwarz term which ensures gauge invariance at one-loop level \cite{Sagnotti:1992qw,Sadov:1996zm} and $\hat \cR$ denotes the curvature two-form in six dimensions. This higher curvature term in F-theory can be understood via its counterpart in M-theory \cite{Antoniadis:1997eg,Bonetti:2011mw,Grimm:2017okk}, as well as from higher-curvature corrections on D7-branes and O7-planes. The latter perspective will be briefly explained in section \ref{sec:solution}.  The field strengths satisfy non-standard Bianchi-identities
\be
\upd \hat G^\ax= \tfr{1}{8}c^\ax\, \tr \hat \cR \w \hat \cR\, ,
\ee 
and the (anti-)-self-duality constraints for the tensors of the tensor- and gravity multiplets, which are imposed at the level of the equations of motion, are collectively given by
\be
 g_{\ax \bx}\,  \hat \ast \, \hat G^\bx=\eta_{\ax \bx} \, \hat G^\bx \, .\lab{selfduality}
\ee
This six-dimensional pseudo-action will be the starting point, similar to the approach to the macroscopic description in \cite{Haghighat:2015ega}.

\subsection{Black string solution in six-dimensional $N=(1,0)$ supergravity}\label{sec:solution}

Two-derivative six-dimensional $N=(1,0)$ supergravity coupled to tensor multiplets has a black string solution which has the same asymptotics as $\mathbb{R} \times \t{S}^1 \times \t{TN}_m$ \cite{Lam:2018jln}. The metric is given by
\begin{equation}\label{metric vacuum solution}
\upd \hat s_{6}^{2}  =  2H^{-1}\upd u\Big(\upd v-\frac{1}{2}H_{5}\upd u\Big)+H\upd s_{4}^{2}\ ,
\end{equation}
with a Taub-NUT metric of (positive) charge $m$, \footnote{Here we choose coordinates for the Taub-NUT space which differ from the ones usually found in the literature.}
\begin{equation}
\upd s_{4}^{2}  =  H_{2}^{-1}m^2\big(\upd \psi+\cos(\theta)\upd \phi\big)^{2}+H_{2}\big(\upd r^{2}+r^{2}\upd \theta^{2}+r^{2}\sin^{2}(\theta)\upd \phi^{2}\big)\ \label{Taub-NUT metric}.
\end{equation}
Furthermore, we have
\begin{equation}
H=\left(\Omega_{\alpha\beta}H_{1}^{\alpha}H_{1}^{\beta}\right)^{1/2}\ ,
\end{equation}
and the harmonic functions on the base $\mathbb{R}^{3}$ of TN${}_m$, denoted by $H_{1}^{\alpha},$
$H_{2}$ and $H_{5}$, are given by
\begin{align}
H_{1}^{\alpha} & =  \mu_{\infty}^{\alpha}+\frac{Q^{\alpha}}{4r}\ ,\qquad
H_{2}  =  m_{\infty}+\frac{m}{r}\ ,\qquad
H_{5}  =  -1+\frac{n}{r}\ . 
\end{align}
We also impose the restriction
\begin{equation}
\Omega_{\alpha\beta}\,\mu_{\infty}^{\alpha}\mu_{\infty}^{\beta}=1\ ,
\end{equation}
in order to get the right asymptotics. The coordinate ranges are given by $0\leq u<\l$ for a length $\l$,
$-\infty< v<\infty,$ $0\leq r<\infty,$ $0\leq\psi<\frac{4\pi}{m},$ $0\leq\theta<\pi$
and $0\leq\phi<2\pi.$ We will use the following dreibein for the $\psi,$ $\phi,$ $\theta$
part of the metric (which we henceforth will refer to as the spherical
part TN$_m^{\mathrm{sph}}$)
\begin{align}
\hat{e}^{1} & =  \sqrt{HH_{2}}\, r\,\big(\sin(\psi)\upd\theta-\cos(\psi)\sin(\theta)\upd\phi\big),\nonumber\\
\hat{e}^{2} & =  \sqrt{HH_{2}}\, r\, \big(\cos(\psi)\upd\theta+\sin(\psi)\sin(\theta)\upd\phi\big),\label{vielbein spherical part}\\
\hat{e}^{3} & =  \sqrt{H/H_{2}}\, m\big(\upd\psi+\cos(\theta)\upd\phi\big)\,.\nonumber 
\end{align}
The near horizon geometry of the metric (\ref{metric vacuum solution}) which is obtained in the limit $r\rightarrow0$ 
is AdS${}_{3}\times \t{S}{}^{3}/\mathbb{Z}_{m}$ with the radius of $ \t{S}{}^{3}/\mathbb{Z}_{m}$ given by $R^2=m \sqrt{\Omega_{\alpha\beta}Q^\alpha Q^\beta}$.
In addition to a non-trivial metric background the solution also requires a radial profile for the scalars $\hat \jmath^{\,\alpha}$  given by \cite{Lam:2018jln}
\begin{equation}
\hat \jmath^{\,\alpha}=\frac{H_{1}^{\alpha}}{H}\, , 
\end{equation}
and non-vanishing three-form backgrounds\footnote{Note that the last term comes with a minus sign since our (anti-)self-duality conventions are opposite to those of \cite{Lam:2018jln}.}
\begin{equation}
\hat {G}^{\alpha}=-\upd v \wedge \upd u\wedge\upd\left(H_{1}^{\alpha}H^{-2}\right)-\star_{4}\upd\left(H_{1}^{\alpha}\right)\, ,\lab{threeforms}
\end{equation}
where $\star_4$ denotes the hodge dual with respect to the Taub-NUT metric $\upd s^2_4$.
We also note that all hypermultiplet scalars are taken to be constant whereas all fermions vanish 
in the background. 

Let us comment on the geometric properties of the Taub-NUT space (\ref{Taub-NUT metric}). 
Firstly, we note that the Taub-NUT space has conical singularities for $m>1$. 
In order to avoid these one can consider multi-centered solutions.  
One can see the metric (\ref{Taub-NUT metric}) for general $m$ as an $m$-centered Taub-NUT space in the limit in which all centers are taken to be coincident. The singularity then arises from the collapsing two-cycles  between the centers of the multi-centered Taub-NUT space. Secondly, we recall that topologically Taub-NUT space is a circle fibration over $\mathbb{R}^3$ and the radius of the circle at infinity is $r_\infty=1/{\sqrt {m_\infty}}$. 
Varying this parameter $r_\infty$ there are two interesting limits which one can consider. The first limit arises when $m_\infty \ll \fr{m}{r}$, i.e.~the NUT circle decompactifies. In this limit the metric \eqref{Taub-NUT metric} approaches (after an additional coordinate transformation) the metric on $\mathbb{R}^4/\mathbb{Z}_m$. In particular, for the case $m=1$ one recovers the black string in flat space. The opposite limit is approached when $m_\infty \gg \fr{m}{r}$. 
This limit is implemented if the circle radius $r_\infty$ is much smaller than the typical length scale of $\mathbb{R}^3$ and leads to  
an effective dimensional reduction of the six-dimensional theory on this circle.

The charges corresponding to the three-forms can be calculated by integrating over the spherical part\begin{equation}
-(2\pi)^2 Q^{\alpha}=\int_{\t{TN}_m^{\rm sph}}\!\!\!\!\hat G^{\alpha}\,
\end{equation}
and are related to the microscopic charges $q^\ax$ via 
\begin{equation}
Q^\ax= q^{\alpha}-\frac{m}{2}c^{\alpha} \,, \lab{charges three-forms in terms of microscopic data}
\end{equation}
as we will demonstrate in the following. Consider type IIB supergravity compactified on the K\"ahler surface $B$, which is the base of the elliptically fibered CY$_3$ in F-theory. Working in conventions $\text{vol}(B)=\frac{1}{2}$ and $\ell_s^2=2\pi$ one can expand the type IIB RR four-form $C_4$ and the K\"ahler form $J_B$ of the base $B$ in harmonic $(1,1)$-forms on $B$
\be
C_4=\hat B^\ax \w \om_\ax\, , \qquad J_B=\hat \jmath^{\, \ax} \om_{\ax}\, , \qquad \text{with }\quad \om_{\ax} \in H^{1,1}(B)\,.
\ee
The two-forms $\hat B^\ax$ are upon dimensional reduction on $B$ identified with the (anti-)self-dual tensors in the six-dimensional gravity- and tensor multiplets, whereas the K\"ahler moduli $\hat \jmath^{\, \ax}$ are interpreted as the scalars in the tensor multiplets. In addition to the bulk type IIB supergravity action there are also localized sources, namely D3-branes, D7-branes and O7-planes, in our setup. The presence of these ten-dimensional localized sources leads to additional six-dimensional couplings, which are crucial for the identification of the macroscopic with the microscopic charges. The D3-brane action contains the standard Chern-Simons action. Now consider $N$ D3-branes with wold-volume $W_{\rm D3}=\Sigma \times C$, where $\Sigma$ is a two-dimensional world-sheet in the six uncompactified dimensions and $C \subset B$ is the curve in the base. Dimensionally reducing the Chern-Simons coupling we obtain
\begin{align}
S_{\rm string}^{\rm CS}&=-\frac{N}{2\pi}\int_{W_{\rm D3}}\!\!\!\!\!\!C_4=-\frac{N}{2\pi}\int_{\Sigma}\hat B^\ax \int_{C} \om_\ax=-\frac{N}{2\pi}\int_{\Sigma} \hat B^\ax q^\bx \int_{B}\om_\ax \w \om_\bx
=-\frac{N}{2\pi} \int_{\Sigma}\eta_{\ax \bx} q^\ax \hat B^\bx \lab{localized}
\end{align}
for the string in six dimensions arising from wrapping the D3-brane over the curve $C$. We obtain further six-dimensional couplings of the two-forms $\hat B^\ax$ by taking into account higher curvature corrections on D7-branes and O7-planes. Expanding again the type IIB four-form $C_4=\hat B^\ax \w \omega_\ax$ and summing over all higher-curvature contributions from D7-branes and O7-planes, as dictated by the F-theory analogue of the D7-brane tadpole cancellation condition (see e.g.~\cite{Grimm:2012yq})
\be
[\text{D7}]+2 [\text{O7}]=12 c_1(B)\, ,
\ee
one obtains the six-dimensional higher-curvature term relevant in the generalized Green-Schwarz mechanism \eqref{6dgeneral}. The total six-dimensional action is then the bulk part \eqref{6dgeneral} coupled to the localized action \eqref{localized}
\be
S^{\sst{(6)}}_{\rm tot}=S^{\sst{(6)}}+S^{\rm CS}_{\rm string}\, .\lab{tot6d}
\ee
Deriving the equations of motion of the (anti-)self-dual tensors we obtain\footnote{We followed footnote 6 in \cite{Giddings:2001yu} and implement the self-duality of the tensors by effectively dividing the source terms in the naive equations of motion derived from \eqref{tot6d} by a factor two.}
\be
\upd \big(g_{\ax \bx}\,\hat \ast \hat G^\bx \big)=(2\pi)^2 \, N\,  \eta_{\ax \bx} q^\bx \, \dx (\Sigma)+\tfr{1}{8} \eta_{\ax \bx} c^\bx \, \tr \hat \cR \w \hat \cR\, .\lab{Beom}
\ee
where $\delta(\Sigma)$ is a four-form delta current localized on the worldsheet of the six-dimensional string. Integrating the resulting equation over TN$_m$ leads to\footnote{The minus sign after the second equality sign is because in our conventions the orientation of TN$^{\mathrm{sph}}_m$ is not the orientation induced by TN$_m$.}
\begin{align}
\fr{1}{(2\pi)^2}\int_{\t{TN}_m}\!\!\!\!\upd \big( g_{\ax \bx}\,\hat \ast \hat G^\bx \big)&=\fr{1}{(2\pi)^2}\eta_{\ax \bx}\int_{\t{TN}_m}\!\!\!\!\upd \hat G^\bx=-\fr{1}{(2\pi)^2}\int_{\t{TN}_m^{\rm sph}}\!\!\!\eta_{\ax \bx}\hat G^\bx=\eta_{\ax \bx}Q^\bx\nn\\
&=N \, \eta_{\ax \bx} q^\bx+\fr{1}{8} \fr{1}{(2\pi)^2}\eta_{\ax \bx} c^\bx \int_{\t{TN}_m}\tr \hat \cR \w \hat \cR\, .
\end{align}
Using furthermore that the first Pontryagin number of Taub-NUT is given by
\be
p_1(\t{TN}_m)=-\fr{1}{2}\fr{1}{(2\pi)^2}\int_{\t{TN}_m}\!\!\!\!\tr \hat \cR \w \hat\cR= 2m\, ,
\ee
we arrive at
\be
Q^\ax=N q^\ax-\fr{m}{2}c^\ax\, ,\lab{qshiftD3}
\ee
which is the desired relation between the macroscopic charge $Q^\ax$ and the microscopic charge $q^\ax$. Most importantly, the classical two-derivative relation $Q^\ax=q^\ax$ obtains a shift proportional to the first Chern class of the base due to the non-trivial topology of the transverse Taub-NUT space.

The relation between the macroscopic and microscopic charges (\ref{charges three-forms in terms of microscopic data}) can also be derived from making contact with the five-dimensional M-theory description. This can be achieved by doing the reduction along the NUT-circle parametrized by $\psi$ to five dimensions \cite{Bonetti:2011mw}. The ansatz for the three-forms is given by 
\begin{equation}
\hat {G}^{\alpha}=G^{\alpha}-F^{\alpha}\wedge\left[m\left(\upd \psi+\cos(\theta)\upd \phi\right)+A^{0}\right],
\end{equation}
where $A^{0}$ is the Kaluza-Klein gauge field and $G^\ax$ is a five-dimensional three-form. The five-dimensional
field strengths are defined in terms of their two-form potentials and the vectors
by
\begin{equation}
G^{\alpha}=\upd B^{\alpha}+A^{\alpha}\wedge F^{0},
\end{equation}
where $F^{0}=\upd A^{0}.$ The three-forms $G^{\alpha}$ are related to
the two-forms $F^{\alpha}=\upd A^{\alpha}$ via the duality relation \eqref{selfduality}.
In order to match the M-theory reduction one has to identify the field strengths $F^{\alpha}$ in terms of the original M-theory field strengths $F_{M}^{\alpha}=\upd A^\ax_M$, $F_{M}^{0}=\upd A_M^0$ via the relation \cite{Bonetti:2011mw}
\begin{equation}
F^{\alpha}=\frac{1}{2}\Big(F_{M}^{\alpha}-\frac{1}{2}c^{\alpha}F_{M}^{0}\Big)\, .\lab{eq:def charges}
\end{equation}
To be more precise, $F^\ax_M$, $F_M^0$ arise from expanding the M-theory four-form field strength as $G_4=F^\ax_M \w \om_\ax+F^0_M \w \om_0$ along vertical divisors. Using (\ref{eq:def charges}) we find
 \begin{equation}
Q^{\alpha}=-\frac{1}{\pi}\int_{\t{S}^{2}}F^{\alpha}=-\frac{1}{{2}\pi}\int_{\t{S}^{2}}\Big(F_{M}^{\alpha}-\frac{1}{2}c^{\alpha}F_{M}^{0}\Big)=q^{\alpha}-\frac{m}{2}c^{\alpha}\, .\lab{chargerelation}
\end{equation}
The relation (\ref{chargerelation}) between the six-dimensional macroscopic charges $Q^\ax$ and the microscopic charges $q^\ax$ is crucial in order to compare our supergravity results with the microscopic data (\ref{central charges}) and (\ref{left level}). Similar shifts have been noticed in related settings \cite{Castro:2007ci,Castro:2007hc,Cano:2018brq}.

\subsection{Classical contributions to central charges and levels}

In the following we will compute the classical contributions to the central charges and levels. By `classical' we mean those contributions which can be obtained from the six-dimensional $(1,0)$ supergravity theory describing our F-theory setup. This is done by extracting coefficients of Chern-Simons terms in three dimensions arising upon dimensionally reducing the six-dimensional action.

\subsubsection{Reduction at asymptotic infinity} The Bekenstein-Hawking entropy of a black hole scales with the area of its event horizon and since this entropy can be calculated from the central charges and levels, one would expect that one has to do the reduction to three dimensions in the near horizon geometry. However, black holes can have hair, in other words degrees of freedom living outside of the horizon and contributing to the microscopic degeneracy \cite{Banerjee:2009uk,Jatkar:2009yd}. A well studied example is provided by considering BMPV black hole \cite{Breckenridge:1996is}, which is microscopically described by a D1-D5 system of type IIB on K3$\times$S$^1$ carrying momentum along S$^1$ and having equal angular momentum in two planes transverse to the D5-brane. Macroscopically this is a five-dimensional rotating black hole. This BMPV black hole can be placed at the center of Taub-NUT to get a four-dimensional black hole, since Taub-NUT space with $m=1$ looks like $\mathbb{R}^4$ in the limit $m_\infty r\ll1$. While the five-dimensional near horizon geometries of the BMPV black hole and its Taub-NUT generalization are the same, the microscopic degeneracies were shown to be different \cite{David:2006yn}. The difference can be explained by invoking the aforementioned hair. For example the center of mass degrees of freedom of the brane system are not captured by the near horizon geometry. Since our setting also includes a Taub-NUT space we expect non-vanishing contributions from hair which has to be taken into account to match the microscopic results of four-dimensional black holes. However, instead of explicitly constructing the hair modes as done in \cite{Banerjee:2009uk} for the BMPV black hole, we use the approach suggested in \cite{Dabholkar:2010rm}. More precisely, we perform the reduction to three dimensions at asymptotic infinity, which for our setting corresponds to sending the dimensionless quantity 
\begin{equation}
r'\equiv \frac{m_\infty r}{m} \rightarrow \infty\,. \label{reduction limit}
\end{equation} Concretely, this means that the reduction is done on the spherical part at large $r'$. According to \cite{Dabholkar:2010rm} the macroscopic levels and central charges, which we will compare with their microscopic counterparts, are then in terms of the asymptotic quantities given by
\begin{align}
k_{L} & =  k_{L}^{\t{asympt}}+\delta_{L}\,,\qquad \qquad
k_{R}  =  k_{R}^{\t{asympt}}+\delta_{R}\, ,\nonumber \\
c_{L} & =  c_{L}^{\t{asympt}}+\Delta\,,\qquad\qquad ~
c_{R}  =  6 k_{R}\, .\label{exterior modes}
\end{align}
The quantities $\delta_{L},$ $\delta_{R}$, $\Delta$ are further $\mathcal{O}(1)$ contributions.  Since the main focus of this work is on the terms that are proportional to the charges of the four-dimensional black hole, i.e.~$q^\ax$ and the NUT charge $m$, we will not compute these contributions. For the terms involving the charges we find that in the classical supergravity reduction the only term leading to different contributions of near horizon and asymptotic geometry is the higher derivative part of the six-dimensional action. Due to our non-trivial transverse geometry this is different from previous work \cite{Dabholkar:2010rm} where also the higher derivative part is the same in the near horizon and asymptotic reductions. In \cite{Dabholkar:2010rm}  the difference of the asymptotic and near horizon reduction manifested itself at the level of $\cO(1)$ contributions, which we do not consider in the following.

The fact that the six-dimensional near horizon geometry does not reproduce the microscopic results for four-dimensional black holes
can also be understood from a different perspective. The microscopic derivations in M-theory have been performed in the 
regime where all volumes of the CY$_3$ are sufficiently large.  This in particular includes the elliptic fiber. The duality to F-theory then implies that we have to consider backgrounds 
on a small NUT circle. Therefore, we expect that the solutions \eqref{metric vacuum solution} can only be used to reproduce the microscopic quantities in the limit (\ref{reduction limit}). Furthermore, the reduced six-dimensional effective action can only be matched to the five-dimensional effective action after adding one-loop corrections coming from the compactification circle. Therefore there is no classical lift of the five-dimensional black string and four-dimensional black hole solution of M-theory to the six-dimensional F-theory solution. The microscopic central charges and levels thus will not just follow from a reduction of six-dimensional supergravity on this background. However, they do follow when one also takes one-loop corrections into account coming from integrating out massive Kaluza-Klein modes on the compact space in the geometry. Calculating these one-loop effects will be the subject of section \ref{sec:quantum}.

\subsubsection{Ansatz for the reduction}

We now present our ansatz for the metric and three-form field strength in order to perform the reduction in the asymptotic geometry, given as a suitable generalization of the ansatz for the near horizon geometry $\ads \times \t{S}^3/\mathbb{Z}_m$. We will do the reduction at an arbitrary radius and compare the asymptotic and near horizon results. 

\paragraph{Near horizon geometry.} The near horizon geometry of the black string solution \eqref{metric vacuum solution} is $\ads \times \t{S}^3/\mathbb{Z}_m$. First consider the  simplest case where $m=1$. This near horizon geometry has an $SO(4)$ isometry group which is identified with rotations on S$^3$. Once perturbations of this background are included, the isometries are gauged and one obtains $SO(4)$ gauge fields. At the level of the algebra, one has $\mathfrak{so}(4)=\mathfrak{su}(2)_L \oplus \mathfrak{su}(2)_R$, such that we effectively have two sets of $\mathfrak{su}(2)$ gauge fields. The ansatz for the dimensional reduction on $\t{AdS}_{3}\times \t{S}^{3}$ can be found in e.g.~\cite{Hansen:2006wu,Dabholkar:2010rm}. We will make use of this ansatz in the following and adapt it appropriately to our setting.

For general NUT-charge $m$ the isometry group $SO(4)$ is broken to ${U}(1)_L/ \mathbb{Z}_m \times {SU}(2)_R$. The unbroken $\mathfrak{u}(1)_L \subset \mathfrak{su}(2)_L$ algebra is generated by the generator $J^3_L$ of the original $\mathfrak{su}(2)_L$ and the total algebra is generated by Killing vectors
\begin{eqnarray}
K_{L} & = & \partial_{\psi},\nonumber \\
K_{R}^{1} & = & \sin(\phi)\partial_{\theta}+\cos(\phi)\cot(\theta)\partial_{\phi}-\frac{\cos(\phi)}{\sin(\theta)}\partial_{\psi},\nonumber \\
K_{R}^{2} & = & \cos(\phi)\partial_{\theta}-\sin(\phi)\cot(\theta)\partial_{\phi}+\frac{\sin(\phi)}{\sin(\theta)}\partial_{\psi},\label{Killing vectors squashed sphere}\\
K_{R}^{3} & = & -\partial_{\phi}\, ,\nonumber 
\end{eqnarray}
which we collectively denote by $K^{i}=\left(K_{R}^{I},K_{L}\right)$ and similarly
$A^{i}=\left(A_{R}^{I},A_{L}\right),$ $F^{i}=\left(F_{R}^{I},F_{L}\right).$ Let us take $\sqrt{\eta_{\ax\bx}Q^\ax Q^\bx}=\frac{1}{m}$ such that the Lens space has unit radius. The appropriate ansatz is \cite{Hansen:2006wu,Dabholkar:2010rm}
\begin{eqnarray}
\upd \hat s_{6}^{2} & = & \upd s_{\t{AdS}_{3}}^{2}+ \delta_{ab}e^{a}e^{b},\label{ansatz metric squashed sphere}\\
\hat{G}^{\alpha} & = & -Q^{\alpha}\left[(2\pi)^2 m\left(e_{3}^{\sst (m)}-\chi_{3}\right)+\ast \t{dvol}(\t{S}^3/\mathbb{Z}_m)\right],\label{ansatz three forms squashed sphere}
\end{eqnarray}
where 
\begin{equation}
e_{3}^{\sst (m)}=\frac{1}{2\pi^{2}}\left[e^{1}\wedge e^{2}\wedge e^{3}-\frac{1}{2}K_{L\,a}e^{a}\wedge F_{L}+\frac{1}{2}K_{R\,a}^{I}e^{a}\wedge F_{R}^{I}\right].\label{e3 for squashed sphere}
\end{equation}
The dreibein is now given by
\begin{equation}
e^{a}=\hat{e}^{a}-K_{L}^{a}A_{L}-K_{R}^{I\,a}A_{R}^{I}\,,
\end{equation}
with $\hat{e}^{a}$ the dreibein (\ref{vielbein spherical part})
in the near horizon limit $r\rightarrow 0$. The three-form $e_{3}^{\sst (m)}$ has the same form as $e_3$, which is used for a reduction on the three-sphere \cite{Hansen:2006wu}, but since $0\leq\psi<\frac{4\pi}{m}$, the integral of $e_{3}^{\sst (m)}$ over the Lens space is given by
\be
\int_{\t{S}^{3}/\mathbb{Z}_{m}}e_{3}^{\sst (m)}=\fr{1}{m}.
\ee
It is also invariant under $U(1)_{L}/\mathbb{Z}_{m}\times SU(2)_{R}$ transformations and one has the relation
\begin{align}
\upd e_{3}^{\sst (m)}&=\frac{1}{16\pi^{2}}F_{L}\wedge F_{L}+\frac{1}{8\pi^{2}} \tr F_{R}\wedge F_{R} \,. 
\end{align}
The three-form $\chi_3$ in the ansatz \eqref{ansatz three forms squashed sphere} is defined by
\be
\chi_3=\fr{1}{16\pi^2} \, A_L \w F_L +\fr{1}{8\pi^2}\, \tr \Big(A_R \w \upd A_R+\fr{2}{3}A_R^3\Big)\, \label{chi3},
\ee
and ensures that the ansatz for the tensors satisfies the Bianchi identity.

\paragraph*{Spherical part of Taub-NUT.}

Consider now a reduction on the spherical part of the metric \eqref{metric vacuum solution} TN$^{\rm sph}_m$ parametrized by $\psi,$ $\phi,$ $\theta$. The Killing vectors of Taub-NUT spacetime are still given by (\ref{Killing vectors squashed sphere})
and form $U(1)_{L}/\mathbb{Z}_{m}\times SU(2)_{R}.$ This implies
that the ansatz of the previous section for the three-forms is still
suitable. The metric of course needs to be adapted and can be taken
as (\ref{ansatz metric squashed sphere}), but now with the vielbein
$\hat{e}^{a}$ of the spherical part of Taub-NUT spacetime (\ref{vielbein spherical part}). The ansatz is thus a straightforward generalization of the one in the near horizon geometry \eqref{ansatz metric squashed sphere}
\begin{eqnarray}
\upd \hat s_{6}^{2} & = & \upd s_{\mathcal{M}_3}^{2}+ \delta_{ab}e^{a}e^{b},\label{ansatz metric squashed sphere2}\\
\hat G^{\alpha} & = & -Q^{\alpha}\left[(2\pi)^2 m\left(e_{3}^{\sst (m)}-\chi_{3}\right)+ \t{dvol}(\mathcal{M}_3)\right]\,,\nn
\end{eqnarray}
with the difference, that we do not take the near horizon limit $r \to 0$ now\footnote{In particular, for the metric ansatz the dreibein $\hat e^a$ is now not taken to be in the $r \to 0$ limit (but for the three-form ansatz it is as defined in the previous paragraph).}. The total metric is therefore TN$_m^{\rm sph}$ fibered over the non-spherical part of the metric, denoted by $\mathcal{M}_3$. We will in the following use this ansatz to calculate the classical parts of the levels and central charges.

\subsubsection{Classical contribution from two- and higher-derivative action}\label{classicalred}

The classical contributions stem from the six-dimensional supergravity action. Aside the leading two-derivative action, also a four-derivative coupling in six dimensions will be of importance to us. We will perform the reduction of the two- and four derivative action separately, and read off their contributions to the levels and central charges from coefficients of three-dimensional Chern-Simons terms.

\paragraph{Two-derivative contribution.}
We calculate the contribution of the two-derivative action to the levels by determining the gauge variation of the reduced action under a $U(1)_L/\mathbb{Z}_m \times SU(2)_R$ gauge transformation. We will do this by integrating the variation of the six-dimensional Lagrangian over the spherical part TN$^{\rm sph}_m$ to obtain the lower dimensional variation. Since $e_3^{\sst (m)}$ is gauge invariant by construction, the only source for a variation under a combined $U(1)_L/\mathbb{Z}_m \times SU(2)_R$ gauge transformation, which is parametrized by $\Lambda$, is $\chi_3$. We therefore obtain\footnote{In our conventions $\int_{\mathcal{M}_6}=\int_{\mathcal{M}_3} \cdot \int_{\mathrm{TN}_m^{\mathrm{sph}}}$.}
\begin{align}
\dx_{\Lambda} \mathscr{L}_{\rm CS} \star_3 1&=-\fr{1}{16 \pi^3}\int_{\t{TN}^{\rm sph}_m} g_{\ax \bx} \, \dx_{\Lambda} \hat G^\ax\w \hat \ast \hat G^\bx=\pi m^2 \eta_{\ax \bx} \, Q^\ax Q^\bx \int_{\t{TN}^{\rm sph}_m} \dx_{\Lambda} \chi_3 \w e_3^{\sst (m)}\nn\\
&=\pi m\, \eta_{\ax \bx} \, Q^\ax Q^\bx \dx_{\Lambda} \chi_3\, , \label{2-der calculation}
\end{align}
where in the second equality we used the (anti-)self-duality condition (\ref{selfduality}).
The lower dimensional variation (\ref{2-der calculation}) is nothing but the gauge variation of a three-dimensional action of the form
\begin{eqnarray}
S_{\mathrm{CS}} & = & \pi m \eta_{\alpha\beta}Q^{\alpha}Q^{\beta}\int_{{\mathcal{M}}_{3}}\chi_{3}\nonumber \\
 & = & \frac{k^{\rm class}_{L}}{8\pi}\int_{\mathcal{M}_3} A_{L}\wedge F_{L}+\frac{k^{\rm class}_{R}}{4\pi}\int_{\mathcal{M}_3}\mathrm{tr}\Big(A_{R}\wedge F_{R}+\frac{2}{3}A_{R}^{3}\Big)\, ,
\end{eqnarray}
with levels and central charges
\begin{align}
k^{\t{2-der}}_{L} & = \frac{1}{2}m\eta_{\alpha\beta}Q^{\alpha}Q^{\beta}=\frac{1}{2}m\eta_{\alpha\beta}\Big(q^{\alpha}-\frac{1}{2}mc^{\alpha}\Big)\Big(q^{\beta}-\frac{1}{2}mc^{\beta}\Big)\,,\nonumber \\
c^{\t{2-der}}_{L}& = c^{\t{2-der}}_{R}=6k^{\t{2-der}}_{R}  =  3m\eta_{\alpha\beta}\Big(q^{\alpha}-\frac{1}{2}mc^{\alpha}\Big)\Big(q^{\beta}-\frac{1}{2}mc^{\beta}\Big)\,.
\end{align}

\paragraph{Higher-derivative contribution.}

In order to find the contribution to the levels and central charges stemming from higher-derivative terms we consider the piece in the six-dimensional action
\be
S^{\sst (6)} \supset \fr{1}{64 \pi^3} \int_{M_6}\eta_{\ax \bx}\, c^\ax\, \hat G^\bx \w \hat  \om^{\rm CS}_{\rm grav}\, ,\lab{6dgrav}
\ee
\noindent where $\hat\om^{\rm CS}_{\rm grav}$ is the gravitational Chern-Simons three-form built of the six-dimensional spin connection. We will compute Chern-Simons terms in three dimensions by integrating \eqref{6dgrav} over the spherical part TN$_m^{\rm sph}$ for general $r$, in particular not taking the near horizon limit. One finds
\begin{align}
\mathscr{L}^{\rm CS}_{\rm h.d.} \star_3 1&=\fr{1}{64 \pi^3}\int_{\t{TN}_m^{\rm sph}}\eta_{\ax \bx} \, c^\ax \, \hat G^\bx \w  \hat \om^{\rm CS}_{\rm grav}\nn \\
&=\fr{1}{16 \pi} \, \eta_{\ax \bx}\, c^\ax \Big(q^\bx-\fr{1}{2}m c^\bx \Big) \bigg[\om_{\rm grav}^{\rm CS} - \fr{1+4r'+2r'^2}{\left(1+r'\right)^4}\, A_L \w F_L\nn\\
&\qquad \qquad \qquad \qquad \qquad \; ~~~ +2 \, \fr{1+4 r'+10 {r'}^2+8 {r'}^3+2 {r'}^4}{(1+r')^4}\, \om^{\rm CS}(A_R) \bigg]\, ,\label{higher derivative result}
\end{align}
where we used $r'=\fr{m_\infty}{m} r$. For the second equality sign in (\ref{higher derivative result}) we only took the parts of $\hat G^\beta$ and $\hat \omega_\mathrm{grav}^\mathrm{CS}$ that lead to Chern-Simons terms in three-dimensions. The choice of dreibein (\ref{vielbein spherical part}) turns out to be very important in order to get proper Chern-Simons terms after reduction. This probably has to do with whether or not the dreibein is globally defined. Details of this calculation can be found in appendix \ref{sec: higher der calculation}. 

Now there are two limits of \eqref{higher derivative result} interesting to us: the near horizon limit $r' \to 0$, where we effectively go to $\ads \times \t{S}^3/ \mathbb{Z}_m$ and the $r' \to \infty$ limit corresponding to performing the 'reduction at infinity'. In the near horizon limit $r' \to 0$ we obtain
\be
 \mathscr{L}^{\rm CS}_{\rm h.d.} \star_3 1=\fr{1}{16 \pi} \eta_{\ax \bx}\, c^\ax \Big(q^\bx-\fr{1}{2}m c^\bx \Big)\, \Big[\om^{\rm CS}_{\rm grav} - \, A_L \w F_L+2 \, \om^{\rm CS}(A_R)\Big]
\ee
from which we read off the following contributions to the central charges and levels
\begin{align}
(c_L-c_R)^{\t{4-der}}&=6\, \eta_{\ax \bx}\, c^\ax \Big(q^\bx-\fr{1}{2}m c^\bx \Big)\, ,\nn\\
k^{\t{4-der}}_R&=\fr{1}{2}\eta_{\ax \bx}\, c^\ax \Big(q^\bx-\fr{1}{2}m c^\bx \Big)\, ,\label{near horizon results}\\
k^{\t{4-der}}_L&=-\fr{1}{2}  \eta_{\ax \bx}\, c^\ax \Big(q^\bx-\fr{1}{2}m c^\bx \Big)\, ,\nn
\end{align}
where we used the fact that the coefficient of the three-dimensional gravitational Chern-Simons term determines the difference between the left- and right-moving central charges. The latter difference can be read off from the gravitational Chern-Simons term by comparing it to
\be
 \mathscr{L}^{\rm CS}_{\rm h.d.} \star_3 1 \supset \frac{c_L-c_R}{96\pi}\, \omega^{\rm CS}_{\rm grav}\, .
\ee
Setting $m=1$ and dropping the charge shift, this is the result obtained in \cite{Dabholkar:2010rm,Haghighat:2015ega}. The shift in the charges is absent in these settings, which involve  black holes in asymptotically flat spacetime, as opposed to our case. We therefore recover prefactors which are in agreement with their results. 

The near horizon results \eqref{near horizon results} turn out not to give the correct classical higher derivative correction to the central charges and levels. In contrast, taking the limit $r' \to \infty$ in \eqref{higher derivative result} one finds 
\be
 \mathscr{L}^{\rm CS}_{\rm h.d.} \star_3 1=\fr{1}{16 \pi} \,   \eta_{\ax \bx}\, c^\ax \Big(q^\bx-\fr{1}{2}m c^\bx \Big) \Big[\om^{\rm CS}_{\rm grav}+4 \om^{\rm CS}(A_R) \Big]\, ,
\ee
such that we obtain
\begin{align}
(c_L-c_R)^{\t{4-der}}&=6 \,   \eta_{\ax \bx}\, c^\ax \Big(q^\bx-\fr{1}{2}m c^\bx \Big)\, ,\nn\\
k_R^{\t{4-der}}&=  \eta_{\ax \bx}\, c^\ax \Big(q^\bx-\fr{1}{2}m c^\bx \Big)\, ,\\
k_L^{\t{4-der}}&=0 \, .\nn
\end{align}
The total classical contributions from the reduction in the asymptotic geometry are therefore given by
\begin{align}
c^{\rm class}_L&=3 m C^2-3 m^2 c_1(B) \cdot C+\fr{3}{4} m^3 c_1(B) ^2+12 c_1(B) \cdot C-6m c_1(B)^2,\nn\\
c^{\rm class}_R&= 6k_R^{\rm class} = 3 m C^2-3 m^2 c_1(B) \cdot C+\fr{3}{4} m^3 c_1(B)^2+6 c_1(B) \cdot C-3m c_1(B)^2,\lab{classcontr}\\
k^{\rm class}_L&=\fr{1}{2}m C^2-\fr{1}{2}m^2 \, c_1(B) \cdot C+\fr{1}{8}m^3 c_1(B)^2.\nn
\end{align}
This is obviously not the full answer, as it does not match the microscopic results (\ref{central charges}) and (\ref{left level}). The mismatch is not surprising because we know that in order to match the six- and five-dimensional effective actions one has to add one-loop corrections to the dimensionally reduced six-dimensional action \cite{Bonetti:2011mw,Bonetti:2013ela}. The results (\ref{classcontr}) are actually equal to the central charges and levels one would find from the five-dimensional action before adding these one-loop corrections. To reproduce the microscopic results one also has to include the one-loop Chern-Simons terms that arise from integrating out the massive Kaluza-Klein modes. This is what we will do in the next section. Adding the classical asymptotic contributions derived in this section to the one-loop induced contributions will lead to a matching of microscopic and macroscopic quantities up to linear order in the charges $(q^\ax, m)$.

\section{Macroscopics in F-theory from 6D: quantum contributions} \lab{sec:quantum}

We now wish to include one-loop Chern-Simons terms in three dimensions and interpret them as additional contributions to the central charges and levels. These loop-induced Chern-Simons terms arise from integrating out massive Kaluza-Klein (KK) modes, which run in the loops of the relevant two-point functions. Since Chern-Simons terms are intimately linked to anomalies in higher dimensions, we anticipate that the relevant three-dimensional fields to be integrated out are KK modes of chiral fields in six dimensions, which can contribute to anomalies. These fields include the six-dimensional gravitino, spin-$\fr{1}{2}$ fermions in the tensor- and hypermultiplets, and the (anti-)self-dual two-forms. Upon reduction to three dimensions these fields lead to massive spin-$\fr{3}{2}$, spin-$\fr{1}{2}$, and three-dimensional chiral vector fields. These somewhat exotic chiral, (anti-)self-dual vector fields in three dimensions were first discussed in \cite{Townsend:1983xs}. One loop corrections due to massive chiral vectors and higher rank tensors were studied in \cite{Bonetti:2012fn}. 

We calculate the loop-induced Chern-Simons terms in the near horizon geometry, but argue that the result is still valid for a reduction at asymptotic infinity. To do the calculation, we first determine the relevant KK-spectrum for our case by truncating the KK-spectrum found in \cite{Deger:1998nm,deBoer:1998kjm} for the case of $N=(2,0)$ supergravity on AdS$_3 \times \t{S}^3$ to the corresponding $N=(1,0)$ spectrum, at least at the two-derivative level. Besides the local Lorentz group representations of the massive fields in three dimensions, we also extract the representations of the fields under the (gauged) $\mathfrak{so}(4)=\mathfrak{su}(2)_L \oplus \mathfrak{su}(2)_R$ isometry of S$^3$, as well as the signs of the three-dimensional masses. We then determine the contribution of a single field for each type to the three-dimensional $\mathfrak{u}(1)_{L}$, $\mathfrak{su}(2)_{R}$ and gravitational Chern-Simons terms. Instead of computing these single field contributions in a direct loop calculation, we make use of the Atiyah-Patodi-Singer (APS) index theorem \cite{Atiyah:1975jf,Atiyah:1976jg,Atiyah:1980jh}. Armed with these results we then sum the contributions over all KK-towers and determine the total contribution employing Zeta-function regularization. In particular, we implement the $\mathbb{Z}_m$ quotient in the sum over KK states. Adding these quantum corrections to the classical ones obtained in section \ref{sec:classical}, we find agreement with the microscopic result up to and including terms of linear order in the charges.

\subsection{Kaluza-Klein spectrum \label{subsec:Kaluza-Klein-spectrum}}

We now determine the $\mathfrak{su}(2)_{L} \oplus \mathfrak{su}(2)_{R}$
representations of the massive spin-$\frac{1}{2}$, spin-$\frac{3}{2}$
and two-form Kaluza-Klein modes before taking the $\mathbb{Z}_m$ quotient.  The
six-dimensional fields that give rise to relevant Kaluza-Klein modes
are the gravitino and self-dual two-form in the gravity multiplet,
the tensorinos and anti-self-dual tensors in the tensor multiplets,
and the hyperinos in the hypermultiplets. The gravitino, tensorinos
and hyperinos are all given by two Weyl fermions subject to a symplectic-Majorana
condition. The tensors obey a reality condition. The $N=(1,0)$ theory
coupled to tensor multiplets can be obtained as a truncation of the
$N=(2,0)$ theory. The spectrum of $N=(2,0)$ supergravity on S$^{3}$
was worked out in \cite{Deger:1998nm,deBoer:1998kjm}. The extra content
we have are the hypermultiplets, but for now we assume that the modes
associated to the fermions in these multiplets fall in the same representations
as the fermions in the tensor multiplets. We now list the massive
modes that one gets without taking into account the symplectic-Majorana
and reality conditions and denote the spectrum in terms of $\mathfrak{so}(4)=\mathfrak{su}(2)_{L}\oplus\mathfrak{su}(2)_{R}$
representations $(j_{L},j_{R})^{\mathrm{sgn}(M)}$ \cite{Deger:1998nm,deBoer:1998kjm}\footnote{In \cite{Deger:1998nm} the spectrum is denoted in terms of the highest
weight vector $(l_{1},l_{2})$ of $\mathfrak{so}(4)$ which is related
to our notation by $l_{1}=j_{L}+j_{R},$ $l_{2}=j_{L}-j_{R}.$},
where $\mathrm{sgn}(M)$ denotes the sign of the mass.
\begin{itemize}
\item Spin-$\frac{3}{2}$:
\[
2\bigoplus_{j_{L}=\frac{1}{2}}^{\infty}\big(j_{L},j_{L}\pm\tfrac{1}{2}\big)^{\mp}.
\]
\item Spin-$\frac{1}{2}$:
\begin{eqnarray*}
2\bigoplus_{j_{L}=\frac{3}{2}}^{\infty}\big(j_{L},j_{L}\pm\tfrac{3}{2}\big)^{\mp} & \oplus & 2\bigoplus_{j_{L}=0}^{1}\big(j_{L},j_{L}+\tfrac{3}{2}\big)^{-}\oplus2\bigoplus_{j_{L}=1}^{\infty}\big(j_{L},j_{L}\pm\tfrac{1}{2}\big)^{\pm}\oplus2\big(\tfrac{1}{2},1\big)^{+}\\
\oplus2\big(0,\tfrac{1}{2}\big)^{+} & \oplus & 2\big(n_{T}+n_{H}\big)\bigoplus_{j_{L}=\frac{1}{2}}^{\infty}\big(j_{L},j_{L}\pm\tfrac{1}{2}\big)^{\pm}\oplus2\left(n_{T}+n_{H}\right)\big(0,\tfrac{1}{2}\big)^{+}.
\end{eqnarray*}
\item Chiral vectors:
\[
\bigoplus_{j_{L}=1}^{\infty}\big(j_{L},j_{L}\pm1\big)^{\mp}\oplus\big(\tfrac{1}{2},\tfrac{3}{2}\big)^{-}\oplus\big(0,1\big)^{-}\oplus n_{T}\bigoplus_{j_{L}=1}^{\infty}\big(j_{L},j_{L}\pm1\big)^{\pm}\oplus n_{T}\big(\tfrac{1}{2},\tfrac{3}{2}\big)^{+}\oplus n_{T}\big(0,1\big)^{+}.
\]
\end{itemize}
The notation we use to denote the representations of the massive KK states is analogous to the notation used in \cite{deBoer:1998kjm}. In particular the notation
\be
\big(j_{L},j_{L}\pm\tfrac{1}{2}\big)^{\mp}=\big(j_{L},j_{L}+\tfrac{1}{2}\big)^{-}\oplus \big(j_{L},j_{L}-\tfrac{1}{2}\big)^{+}
\ee
is a shorthand notation for the existence of two infinite towers of KK modes in the spectra listed above. We furthermore want to mention that the three-dimensional fermions are Dirac spinors and the chiral vectors are complex. 

Applying the symplectic-Majorana and reality conditions for the gravitino and the tensors means that
modes with quantum numbers $j_{L}^{3},$ $j_{R}^{3}$ are mapped to
modes with quantum numbers $-j_{L}^{3},$ $-j_{R}^{3}$ \cite{Ishiki:2006rt}.
Here $j_{L}^{3},$ $j_{R}^{3}$ are the eigenvalues of the generators
of $\mathfrak{u}(1)_{L}\subset\mathfrak{su}(2)_{L}$ and $\mathfrak{u}(1)_{R}\subset\mathfrak{su}(2)_{R}$
respectively. This effectively means that we only have to sum over modes
with $j_{L}^{3}\geq0$. The Kaluza-Klein spectrum for a reduction
on S$^{3}/\mathbb{Z}_{m}$ can then be obtained by projecting onto the $\mathbb{Z}_{m}$-invariant
states of the spectrum on S$^3$, as shown above. This means that we only keep
those states which have $j_{L}^{3}=\frac{1}{2}mk$ for some $k\in\mathbb{Z}_{\geq0}$
\cite{Ishiki:2006rt}. 

If one performs the reduction in the asymptotic geometry one reduces on a squashed Lens space, where the radius of the two-sphere inside the squashed three-dimensional geometry is taken to be large. We expect that the representation content of the KK spectrum does not get altered by the squashing. Note that, due to the asymptotic NUT circle, the masses of the Kaluza-Klein modes remain finite. In addition we assume, that the squashing of the S$^3/\mathbb{Z}_m$ does not change the sign of the mass of the KK states. These assumptions essentially imply, that we can do the loop computation in the near horizon geometry and use the spectrum on S$^3/\mathbb{Z}_m$.

\subsection{One-loop Chern-Simons terms from KK spectrum \label{sec:Calculation-of-one-loop}}

Quantum corrections to Chern-Simons
terms can be interpreted as compensations for the parity violation
introduced by families of massive fields, after they are integrated
out \cite{Bonetti:2013ela}. The fields that contribute in our case to the three-dimensional parity
anomaly are massive spin-$\frac{1}{2}$ fermions, spin-$\frac{3}{2}$
fermions and massive vectors in three dimensions. We can thus calculate these corrections
by calculation of the parity-violating piece of the effective action
which can be expressed using the Atiyah-Patodi-Singer $\eta-$invariant
\cite{AlvarezGaume:1984nf} corresponding to the relevant Dirac operator.
This $\eta-$invariant can be expressed in Chern-Simons terms by extending
the Dirac operator to one dimension higher and using the
Atiyah-Patodi-Singer index theorem \cite{AlvarezGaume:1984nf}. This
calculation is valid for three-dimensional Riemannian manifolds of
the form $\mathcal{M}_{3}=\mathbb{R}\times\mathcal{M}_{2},$ where
$\mathcal{M}_{2}$ is a compact manifold without boundary. Since we
are doing the reduction at infinity, where the three-dimensional manifold
(after Wick rotation) is of the form $\mathbb{R}^{2}\times \mathrm{S}^{1}$,
the index theorem is indeed applicable by treating this manifold as
$\mathbb{R}\times \mathrm{S}_{R}^{1}\times \mathrm{S}^{1},$ where we take the radius
of the S$_{R}^{1}$ circle to be very large.

We now first treat the spin-$\frac{1}{2}$ fermions, the spin-$\frac{3}{2}$ fermions and the massive vectors separately. The loop corrections induced by these three types of fields are listed in table \ref{looptable}. After these corrections are determined, we sum the latter
over the spectrum determined in the previous subsection to compute
the full one-loop correction to the central charges and levels.

\paragraph*{Spin-$\boldsymbol{\frac{1}{2}}$ fermions.}

We consider a massive spin-$\frac{1}{2}$ fermion coupled to the gauge
fields $A=(A_{L},\ A_{R})$ taking values in the Lie algebra $\mathfrak{u}(1)_{L}\oplus\mathfrak{su}(2)_{R}$
and to an external gravitational field denoted by the vielbein $e$ with spin connection $\omega$.
The parity anomaly resulting from this particle can be canceled by a term \cite{AlvarezGaume:1984nf}
\begin{equation}
-i\pi\, \mathrm{sgn}(M)\int_{\mathcal{M}_{3}}Q_{\frac{1}{2}}(A,\omega),\label{counterterm spin 1/2 6 to 5-1}
\end{equation}
with
\begin{equation}
\upd Q_{\frac{1}{2}}(A,\omega)=\hat{A}(\mathcal{M}_{3})\wedge\mathrm{ch}(F_{L})\wedge\mathrm{ch}(F_{R})\big|\, ,\lab{dQ}
\end{equation}
where the vertical dash denotes that we pick out the four-form contribution of the whole expansion on the right hand side of \eqref{dQ}. The form at the right-hand side is the index density appropriate to
the Dirac operator for spin-$\frac{1}{2}$ particles in four dimensions. It is expressed in terms of the Dirac genus $\hat A$ and Chern character, which
have an expansion
\begin{eqnarray}
\hat{A}(\mathcal{M}_{3}) & = & 1+\frac{1}{\left(4\pi\right)^{2}}\frac{1}{12} \tr \cR\wedge \cR +...\ ,\\
\mathrm{ch}(F) & = & r+\frac{i}{2\pi} \tr F -\frac{1}{2}\frac{1}{\left(2\pi\right)^{2}}\tr F\wedge F - \frac{i}{6}\frac{1}{\left(2\pi\right)^{3}} \tr F\wedge F\wedge F +...\ ,\nonumber 
\end{eqnarray}
where $r$ is the dimension of the representation of the gauge group, under which the spin-$\fr{1}{2}$ fermion transforms. We use that
\begin{equation}
\hat{A}(\mathcal{M}_{3})\wedge\mathrm{ch}(F_{L})\wedge\mathrm{ch}(F_{R})\big|=\frac{1}{\left(2\pi\right)^{2}}\Big(\,\frac{r}{4}F_{L}\wedge F_{L}-\frac{1}{2}\tr F_{R}\wedge F_{R} +\frac{r}{48} \tr \cR\wedge \cR\,\Big)\, ,
\end{equation}
where now $r$ is the dimension of the $\mathfrak{su}(2)_{R}$ representation
of the spin-$\frac{1}{2}$ fermion and we used that the generator
for the $\mathfrak{u}(1)_{L}$ is given in terms of the Pauli matrices
by $-\frac{i}{2}\sigma_{3}$. We find that the counterterm to cancel the parity anomaly is then
given by
\begin{equation}
\mathrm{sign}(M)\left(-\frac{ir}{16\pi}A_{L}\wedge F_{L}+\frac{i}{8\pi}\omega_{\mathrm{CS}}\left(A_{R}\right)-\frac{ir}{192\pi}\omega_{\mathrm{grav}}^{\mathrm{CS}}\right).
\end{equation}
Note that these are the corrections to the action on the Riemannian
manifold. We still have to Wick rotate to Lorentzian signature by multiplying
with a factor $i$, which yields the counter terms
\begin{equation}
\mathrm{sign}(M)\left(\frac{r}{16\pi}A_{L}\wedge F_{L}-\frac{1}{8\pi}\omega_{\mathrm{CS}}\left(A_{R}\right)+\frac{r}{192\pi}\omega_{\mathrm{grav}}^{\mathrm{CS}}\right).
\end{equation}

\paragraph*{Spin-$\boldsymbol{\frac{3}{2}}$ fermions.}

For spin-$\frac{3}{2}$ fermions the counterterm is given by (\ref{counterterm spin 1/2 6 to 5-1})
with \cite{AlvarezGaume:1984nf,AlvarezGaume:1984dr}
\begin{align}
\upd Q_{\frac{3}{2}}(A,\omega) & = \hat{A}(\mathcal{M}_{3})\wedge \left[ \mathrm{tr}\,\exp \bigg({\frac{i\cR}{2(2\pi)^2}}\bigg)-1\right]\wedge\mathrm{ch}(F_L) \w \mathrm{ch}(F_R)\Big|\,.
\end{align}
Using that $\mathrm{tr}\,\exp \left( \frac{i \cR}{2\pi}\right)-1=3- \frac{1}{2(2\pi)^2}\mathrm{tr}\,\cR \w \cR+ ...,$
we find that
\begin{align}
\upd Q_{\frac{3}{2}}(A, \omega) & =\frac{3r}{4\left(2\pi\right)^{2}}F_{L}\wedge F_{L}-\frac{3}{2\left(2\pi\right)^{2}} \tr F_{R}\wedge F_{R} - \frac{7r}{16 (2\pi)^2} \tr \cR\wedge \cR \,.
\end{align}
The counterterm to the Lorentzian action then becomes
\begin{equation}
\mathrm{sign}(M)\left(\frac{3r}{16\pi}A_{L}\wedge F_{L}-\frac{3}{8\pi}\omega_{\mathrm{CS}}\left(A_{R}\right)-\frac{7r}{64 \pi}\omega_{\mathrm{grav}}^{\mathrm{CS}}\right).
\end{equation}

\paragraph*{Chiral vectors.}

In this case we were unaware of the existence of an appropriate index theorem in the literature.
When ignoring the gauge fields one gets \cite{AlvarezGaume:1984dr}
\begin{equation}
\mathrm{ind}\,iD_{A}=\frac{1}{2}\int_{M}L(M)|,\label{eq:index tensor}
\end{equation}
where the Hirzebruch $L$-polynomial is given by
\begin{eqnarray}
L(M) & = & 1-\frac{1}{\left(2\pi\right)^{2}}\frac{1}{6} \tr \cR\wedge \cR + ...\ .
\end{eqnarray}
The equality in (\ref{eq:index tensor}) only holds for the four-form and we multiplied the right hand side by two with respect to the result in \cite{AlvarezGaume:1984dr} since we consider complex instead of real vector fields. However,
we now use that the $L-$polynomial according to the Hirzebruch signature
theorem also determines the Hirzebruch signature:
\begin{equation}
\tau=\int_{M}L(M)|.
\end{equation}
If one instead considers a tensor product with another vector bundle
the Hirzebruch theorem becomes \cite{Eguchi:1980jx}
\begin{equation}
\tau=\int_{M}L(M)\wedge\mathrm{ch}\left(2F\right)|.
\end{equation}
Based on these considerations, we now postulate that
\begin{equation}
\mathrm{ind}\,iD_{A}=\frac{1}{2}\int_{M}L(M)\wedge\mathrm{ch}\left(2F\right)|.\label{index self-dual tensor}
\end{equation}
In \cite{Bonetti:2013ela} one-loop corrections are computed that
one gets when integrating out massive chiral Kaluza-Klein modes after the reduction
from six to five dimensions on a circle. The authors do this by explicit calculation of the
diagrams. In appendix \ref{sec:6D-to-5D} we reproduce these results using the index theorems
in which we also use the index (\ref{index self-dual tensor}). This
is some non-trivial evidence that this is the right quantity. 

Using the index (\ref{index self-dual tensor}) the counterterm is now given by
\begin{equation}
i\pi\,\mathrm{sign}(M)\int_{\mathcal{M}_{3}}Q_{\rm vec}(A,\omega)\label{counterterm self-dual}
\end{equation}
where
\begin{eqnarray}
\upd Q_{\rm vec}(A,\omega) & = & \frac{1}{2}L(M)\wedge\mathrm{ch}\left(2F_{L}\right)\wedge\mathrm{ch}\left(2F_{L}\right)|.
\end{eqnarray}
Notice that (\ref{counterterm self-dual}) has an extra minus-sign
with respect to (\ref{counterterm spin 1/2 6 to 5-1}) which is
caused by the vectors being bosons \cite{AlvarezGaume:1984dr}. We
then find
\begin{align}
\upd Q_{\rm vec}(A, \omega)& =  \frac{r}{2\left(2\pi\right)^{2}}F_{L}\wedge F_{L}-\frac{1}{\left(2\pi\right)^{2}} \tr F_{R}\wedge F_{R} - \frac{1}{12}\frac{r}{\left(2\pi\right)^{2}} \tr \cR\wedge \cR \,.
\end{align}
This implies that the counterterms to the Lorentzian action are given
by
\begin{equation}
\mathrm{sign}(M)\left(-\frac{r}{8\pi}A_{L}\wedge F_{L}+\frac{1}{4\pi}\omega_{\mathrm{CS}}\left(A_{R}\right)+\frac{r}{48\pi}\omega_{\mathrm{grav}}^{\mathrm{CS}}\right).
\end{equation}

\paragraph*{Corrections to the levels and central charges.}

Note that all the corrections above were derived for an arbitrary
representation under $\mathfrak{u}(1)_{L}\oplus\mathfrak{su}(2)_{R}$
specified by the quantum numbers $j_{L}^{3}$ and $j_{R}$. Expressing
the left Chern-Simons term in the representation we used in the classical
part gives a factor $2\left(j_{L}^{3}\right)^{2}$ and expressing
the right Chern-Simons terms in the fundamental representation gives
a factor $\frac{2}{3}j_{R}\left(j_{R}+1\right)\left(2j_{R}+1\right).$
We also use that the dimension of the representation under $\mathfrak{su}(2)_{R}$
is given by $2j_{R}+1.$ The constants $\alpha_{L},$ $\alpha_{R},$
$\alpha_{\mathrm{grav}}$ in front of the Chern-Simons terms $\omega_{\mathrm{CS}}^{\mathrm{}}(A_{L}),$
$\omega_{\mathrm{CS}}^{\mathrm{}}\left(A_{R}\right)$ and $\omega_{\mathrm{grav}}^{\mathrm{CS}}$
are then given in table \ref{looptable}.

\begin{table}[h]
\centering
{\renewcommand{\arraystretch}{1.6}
\begin{tabular}{c|c|c|c}
\Xhline{2\arrayrulewidth}
 & spin-$\frac{1}{2}$  & spin-$\frac{3}{2}$  &  chiral vectors\\[0.1cm]
\hline 
\hline
$\alpha_{L}$ & $\frac{1}{2}(j^3_{L})^{2}\left(2j_{R}+1\right)$ & $\frac{3}{2}(j^3_{L})^{2}\left(2j_{R}+1\right)$ & $-(j^3_{L})^{2}\left(2j_{R}+1\right)$\\[0.1cm]

$\alpha_{R}$ & $-\frac{1}{3}j_{R}\left(j_{R}+1\right)\left(2j_{R}+1\right)$ & $-j_{R}\left(j_{R}+1\right)\left(2j_{R}+1\right)$ & $\frac{2}{3}j_{R}\left(j_{R}+1\right)\left(2j_{R}+1\right)$\tabularnewline
 
$\alpha_{\mathrm{grav}}$ & $\frac{1}{48}\left(2j_{R}+1\right)$ & $-\frac{7}{16}\left(2j_{R}+1\right)$ & $\frac{1}{12}\left(2j_{R}+1\right)$\tabularnewline\Xhline{2\arrayrulewidth}
 
\end{tabular} 
}
\caption{Contributions of a single field to the left-, right- and gravitational Chern-Simons terms. The table should be read as $\ax_I= \fr{\t{sgn}(M)}{4 \pi} \times$(entry of table).}
\lab{looptable}
\end{table}
We now sum the contributions of table \ref{looptable}
over the spectrum determined in section \ref{subsec:Kaluza-Klein-spectrum}.
Since the projection condition is $j_{L}^{3}=\frac{1}{2}mk$ for $k\in\mathbb{Z}_{\geq 0},$
we first sum over all representations which
contain a state with $j_{L}^{3}=\frac{1}{2}mk.$ These are just the
representations labeled by $j_{L}=\frac{1}{2}mk,$ $\frac{1}{2}mk+1,$
$...$. Finally, we sum over all $k\in\mathbb{Z}_{\geq 0}.$ The sums
we encounter are of the form $\sum_{n}f(n)$ where $n$ runs over
integers or half integers. We regularize the infinite, divergent sums using zeta-function regularization.
In particular, we use the regularized sums
\begin{eqnarray}
 \sum_{n=1}^{\infty}& 1  =  -\frac{1}{2},\ \ \ \ & \sum_{n=1}^{\infty}n=-\frac{1}{12},\nonumber \\
 \sum_{n=1}^{\infty}& n^{2}  =  0,\ \ \ \ \quad  &\sum_{n=1}^{\infty}n^{3}=\frac{1}{120}.\label{zeta function regularization}
\end{eqnarray}
It is worth noting that regularizing the contributions arising from integrating out infinitely many massive modes is in general very subtle.\footnote{See \cite{Grimm:2018ohb} for a recent detailed discussion.} Firstly, applying zeta-function regularization is only possible if the higher-dimensional theory is anomaly free \cite{Corvilain:2017luj}. Secondly, in a theory with gravity, one expects that there is a UV cut-off in the 
lower-dimensional theory set by the scale at which gravity becomes strongly coupled \cite{ArkaniHamed:2005yv,Dvali:2007hz}. It turns out that the result from this regularization scheme agrees with the zeta-function regularization.  

When calculating the one-loop corrections, we also make use of the identities
\begin{eqnarray}
n_{H} & = & 273-29n_{T},\nn\\
n_{T} & = & h^{1,1}(B)-1=9-c_{1}(B)^2,\label{anomaly and calabi-yau identity} 
\end{eqnarray}
where the first one is the anomaly cancellation condition in 6D. The relations between the coefficients $\ax_I$, which we compute by performing the infinite sums over Kaluza-Klein states, and the levels and central charges is given by
\be
k_L=8\pi  \ax_L\, ,\qquad k_R=4\pi \ax_R\, \, \qquad c_L-c_R=96\pi \ax_{\rm grav}\, .
\ee
The explicit calculation of the sums can be found in appendix \ref{sec:Summation-of-6D}, but the results are given by
\begin{eqnarray}
\Delta k_{L}^{\mathrm{loop}} & = & -\frac{m^{3}}{8}c_{1}(B)^{2},\nn \\
\Delta k_{R}^{\mathrm{loop}} & = & \frac{m^{3}}{24}c_{1}(B)^{2}+\frac{m}{3}c_{1}(B)^{2}+m\,,\lab{loopcontr} \\
\Delta\left(c_{L}-c_{R}\right)^{\mathrm{loop}} & = & 6m+2m\,c_{1}(B)^{2}\,,\nn
\end{eqnarray}
up to terms of $\cO(1)$ which are independent of the charges $(q^\alpha, m)$. 

We notice that the one loop corrections to $\Delta k_{L}^{\mathrm{loop}}$ and $\Delta\left(c_{L}-c_{R}\right)^{\mathrm{loop}}$ differ for the two cases $m=1$ and $m>1$ at the level of the constants.
This difference would disappear when adding 4 $(\tfrac{1}{2},0)^-$ representations to the spin-$\frac{1}{2}$ spectrum. Since we are however interested in contributions scaling with the charges we do not comment further on this case distinction.

\section{Summary and 4D/5D correspondence} \lab{sec: comp with earlier work}

We derived the central charges $c_{L, R}$ and levels $k_{L, R}$ of a $(0,4)$ SCFT
corresponding to an F-theory geometry $\mathbb{R}\times \mathrm{S}^{1}\times \mathrm{TN}_{m}\times \mathrm{CY}_{3}$ with a D3 brane wrapped around $\mathrm{S}^{1}\times C,$ from six-dimensional $(1,0)$ supergravity. Combining the classical contributions \eqref{classcontr} with the one loop results \eqref{loopcontr} leads to the total result
\begin{align}
c_{L} & =  3mC^{2}-3m^{2}c_{1}(B)\cdot C+m^{3}c_{1}(B)^{2}+12c_{1}(B)\cdot C+12m-2mc_{1}(B)^{2}\,,\nonumber \\
c_{R} & =  6k_{R}=3mC^{2}-3m^{2}c_{1}(B)\cdot C+m^{3}c_{1}(B)^{2}+6c_{1}(B)\cdot C+6m-mc_{1}(B)^{2}\,,\label{complete expressions levels and charges}\nonumber  \\
k_{L} & =  \frac{1}{2}mC^{2}-\frac{1}{2}m^{2}c_{1}(B)\cdot C\, ,
\end{align}
for the central charges, again up to $\cO(1)$ contributions independent of the charges. We reproduced (\ref{central charges}) up to constants and we reproduced (\ref{left level}) exactly.

\paragraph*{The 4D/5D correspondence.}
As already mentioned in the introduction, the setting we studied in this paper compactified to four dimensions is the four-dimensional side of the 4D/5D correspondence \cite{Gaiotto:2005gf,Behrndt:2005he}. The five-dimensional side of this correspondence is given by a five-dimensional black hole with flat asymptotics which can be uplifted to the asymptotically flat string in six dimensions. Therefore, it is interesting to compare the results (\ref{complete expressions levels and charges}) with the macroscopic derivation of the central charges and levels of the asymptotically flat black string. The near horizon geometry of this string is the same as in our case with NUT-charge $m=1.$ The asymptotic geometry is however different in both cases. Microscopically the asymptotically flat black string was studied in  \cite{Haghighat:2015ega} and is described by an F-theory geometry $\mathbb{R}^{1,4}\times \mathrm{S}^{1}\times \mathrm{CY}_{3}$ with a D3-brane wrapping S$^1 \times C$ for a curve $C\subset B$ in the base of the Calabi-Yau threefold. It was found that microscopically, after performing a topological duality twist \cite{Martucci:2014ema}, the effective two-dimensional theory of the D3-brane wrapping the curve preserves $N=(0,4)$ supersymmetry and has left- and right moving $SU(2)_{L, R}$ current algebras. A similar analysis was carried out in \cite{Lawrie:2016axq}, where various topological duality twists of the D3-brane worldvolume theory preserving $N=(0, n)$ with $n=2,4,6,8$ supersymmetry are studied. A microscopic count of degrees of freedom of the worldvolume theory leads to\footnote{The center of mass contributions are subtracted.} 
\begin{align}
c_L^{\t{flat}}&=3 C^2+9 c_1(B) \cdot C+2\, ,\nn\\
c_R^{\t{flat}}&=6 \, k_R^{\t{flat}}=3 C^2 +3 c_1(B) \cdot C\, ,\label{microscopic data asympt flat black string}\\
k_L^{\t{flat}}&=\fr{1}{2} C^2-\fr{1}{2} c_1(B) \cdot C\, \nonumber. 
\end{align}

The central charges and levels  (\ref{microscopic data asympt flat black string}) corresponding to the asymptotically flat black string can be, up to the constants, reproduced from six-dimensional (1,0) supergravity \cite{Haghighat:2015ega}. 
Ignoring the constants, the expressions (\ref{complete expressions levels and charges}) found in our setting reduce for $m=1$ to (\ref{microscopic data asympt flat black string}). This nicely fits in the picture of the 4D/5D correspondence \cite{Gaiotto:2005gf,Behrndt:2005he}. However, the subleading terms in (\ref{microscopic data asympt flat black string})
that for the asymptotically flat case are higher-derivative contributions, come in our case
both from the two-derivative and the higher derivative part of the
action.

\section{Discussion} \lab{sec: conclusion}

In this work we considered the F-theory geometry $\mathbb{R} \times \mathrm{S}^1 \times \mathrm{TN}_m \times \mathrm{CY}_3$ with a D3-brane wrapped around S$^1 \times C$, where $C \subset B$ is a curve in the base of the elliptically fibered Calabi-Yau threefold. From a six-dimensional supergravity perspective we reproduced the central charges and levels of the CFT corresponding to the long wavelength limit of this set-up, which were calculated from a microscopic viewpoint via M-theory in \cite{Bena:2006qm}. Macroscopically the levels and central charges are determined by coefficients of gauge and gravitational Chern-Simons terms in the three-dimensional effective action obtained after reduction on the spherical part of the six-dimensional space-time. We identify and include a shift in the identification of the macroscopic with the microscopic charges caused by the Green-Schwarz term in the pseudo-action. Performing the reduction at asymptotic infinity and the inclusion of one-loop Chern-Simons terms arising from integrating out massive Kaluza-Klein modes are the crucial ingredients to obtain this non-trivial matching.

An alternative way to reproduce the central charges and levels is by employing the effective action obtained by reducing M-theory on CY$_3$.  Upon further reduction on the near horizon geometry AdS$_3 \times \text{S}^2$ one can generate Chern-Simons terms, which in turn determine the central charges and right level \cite{Kraus:2005vz}. The right moving current algebra is identified with the $SU(2) $ isometry group of the sphere. Note that this $SU(2)$ is not the same as $SU(2)_R$ in our six-dimensional picture. Reproducing the left level from five dimensions works in a slightly different way, since the $U(1)_L/\mathbb{Z}_m$ corresponds to the circle of Taub-NUT that is now hidden in the geometry. However, one can do it using the data of the effective five-dimensional action in a similar way as we did in section \ref{sec:left level} for the microscopic calculation. Although this was not our main motivation, in the six-dimensional approach the geometric realization of the left moving current algebra is more clear. 

Our results provide a first step for embedding and studying four-dimensional black holes in F-theory from a macroscopic point of view. Since F-theory has proven to be a particularly successful framework for particle physics model building, its (quantum) gravitational aspects, which remained mostly unexplored up to now, might reveal interesting physics as well. We believe that our work provides a natural next step in the development and study of black holes in F-theory.

An obvious and interesting generalization of this work would be to include vector multiplets in the six-dimensional $N=(1,0)$ supergravity theory we take as our starting point. These are realized in F-theory compactifications on elliptically fibered Calabi-Yau manifolds in which fiber degenerations do not leave the total space of the fibration smooth. To the best of our knowledge there is no microscopic prediction for this case available in the literature. In addition, since the one-loop corrections played such an important role, it would be interesting to understand the relation between six- and five-dimensional supergravity solutions better.

\paragraph*{Acknowledgements}

We  thank Babak Haghighat and Jo\~ao Gomes for initial collaboration on this project. It is also a pleasure to thank Atish Dabholkar, Dario Martelli, Miguel Montero, Sameer Murthy, Sakura Sch\"afer-Nameki and Phil Szepietowski for valuable discussions and correspondence. 

This work was supported in part by the D-ITP consortium, a program of the Netherlands Organization for Scientific Research (NWO) that is funded by the Dutch Ministry of Education, Culture and Science (OCW), and by the NWO Graduate Programme.

\appendix

\section{6D to 3D reduction higher derivative term \label{sec: higher der calculation}}

In this appendix we give some more details of the reduction of the
six-dimensional higher derivative term to three dimensions. In particular,
we calculate the part of the integral
\begin{equation}
\int_{\mathrm{TN}_{m}^{\mathrm{sph}}}\hat{G}^{\alpha}\wedge\hat{\omega}_{\mathrm{grav}}^{\mathrm{CS}}
\end{equation}
that leads to three-dimensional Chern-Simons terms. In order to do
the reduction we first decompose the spin connection corresponding
to the ansatz (\ref{ansatz metric squashed sphere2}) to determine
the parts that lead to Chern-Simons terms in three dimensions.
Denoting indices of the non-spherical part $\mathcal{M}_{3}$ of the
black string solution by $\tilde{a}=1,2,3$ and a vielbein of $\mathcal{M}_{3}$
by $\hat{e}^{\tilde{a}},$ the spin connection $\omega$ with respect
to the vielbein of the ansatz, $e^{\tilde{a}}\equiv\hat{e}^{\tilde{a}},$
$e^{a}$, can be expressed as \cite{Duff:1986hr}
\begin{eqnarray}
\omega_{\tilde{a}\tilde{b}} & = & \hat{\omega}_{\tilde{a}\tilde{b}}+\frac{1}{2}F_{\tilde{a}\tilde{b}}^{i}K_{c}^{i}e^{c},\nonumber \\
\omega_{\tilde{a}b} & = & \frac{1}{2}F_{\tilde{a}\tilde{c}}^{i}K_{b}^{i}\hat{e}^{\tilde{c}},\\
\omega_{ab} & = & \hat{\omega}_{ab}+\left(\hat{\nabla}_{a}K_{b}^{i}\right)A^{i}.\nonumber 
\end{eqnarray}
Here $\hat{\omega}_{\tilde{a}\tilde{b}}$ are the components of the
spin connection $\hat{\omega}_{\mathcal{M}_{3}}$ with respect to
the vielbein $\hat{e}^{\tilde{a}}$ of $\mathcal{M}_{3}$ and $\hat{\omega}_{ab}$
are the components of the spin connection $\hat{\omega}_{\mathrm{sph}}$
with respect to the vielbein $\hat{e}^{a}$ of the spherical part
of the black string solution. From the expression of the gravitational
Chern-Simons term,
\begin{equation}
\hat{\omega}_{\mathrm{grav}}^{\mathrm{CS}}= \tr \left(\omega\wedge\mathrm{d}\omega+\frac{2}{3}\omega^{3}\right),
\end{equation}
it is immediately clear that if we are interested in three-dimensional
Chern-Simons terms, we can restrict to 
\begin{eqnarray}
\omega_{\tilde{a}\tilde{b}} & = & \hat{\omega}_{\tilde{a}\tilde{b}},\nonumber \\
\omega_{\tilde{a}b} & = & 0,\\
\omega_{ab} & = & \hat{\omega}_{ab}+\left(\hat{\nabla}_{a}K_{b}^{i}\right)A^{i}.\nonumber 
\end{eqnarray}
This is a direct sum connection, hence
\begin{equation}
\hat{\omega}_{\mathrm{grav}}^{\mathrm{CS}}=\omega^{\mathrm{CS}}\left(\hat{\omega}_{\mathcal{M}_{3}}\right)+\omega^{\mathrm{CS}}\left(X\right),
\end{equation}
where $X$ is a connection with components $\hat{\omega}_{ab}+\left(\hat{\nabla}_{a}K_{b}^{i}\right)A^{i}$
and
\begin{eqnarray}
\omega^{\mathrm{CS}}\left(\hat{\omega}_{\mathcal{M}_{3}}\right) & \equiv & \tr \left(\hat{\omega}_{\mathcal{M}_{3}}\wedge\mathrm{d}\hat{\omega}_{\mathcal{M}_{3}}+\frac{2}{3}\hat{\omega}_{\mathcal{M}_{3}}^{3}\right),\nonumber \\
\omega^{\mathrm{CS}}\left(X\right) & \equiv & \tr \left(X\wedge\mathrm{d}X+\frac{2}{3}X^{3}\right).
\end{eqnarray}
Notice that $\omega^{\mathrm{CS}}\left(\hat{\omega}_{\mathcal{M}_{3}}\right)=\omega_{\mathrm{grav}}^{\mathrm{CS}}$
is the gravitational Chern-Simons term of $\mathcal{M}_{3}$. 

The only part of $\hat{G}^{\alpha}$ relevant for Chern-Simons terms
in three dimensions is 
\begin{equation}
-Q^{\alpha}\left(2\pi\right)^{2}m\left(e_{3}^{(m)}-\chi_{3}\right).
\end{equation}
Here $\chi_{3}$ has all its legs on $\mathcal{M}_{3}$ which means
that its wedge product with $\hat{\omega}_{\mathrm{grav}}^{\mathrm{CS}}$
only gets a contribution of $\omega^{\mathrm{CS}}\left(\hat{\omega}_{\mathrm{sph}}\right).$
We can then expand
\begin{eqnarray}
\int_{\mathrm{TN}_{m}^{\mathrm{sph}}}\hat{G}^{\alpha}\wedge\hat{\omega}_{\mathrm{grav}}^{\mathrm{CS}} & = & -Q^{\alpha}\left(2\pi\right)^{2}m\int_{\mathrm{TN}_{m}^{\mathrm{sph}}}\left[e_{3}^{(m)}\wedge\omega_{\mathrm{grav}}^{\mathrm{CS}}+e_{3}^{(m)}\wedge\omega^{\mathrm{CS}}\left(X\right)-\chi_{3}\wedge\omega^{\mathrm{CS}}\left(\hat{\omega}_{\mathrm{sph}}\right)\right].\nonumber \\
\label{eq:integral appendix}
\end{eqnarray}
The separate integrals are given by\footnote{We used Mathematica to calculate the second and third integral.}
\begin{eqnarray}
\int_{\mathrm{TN}_{m}^{\mathrm{sph}}}e_{3}^{(m)}\wedge\omega_{\mathrm{grav}}^{\mathrm{CS}} & = & -\frac{1}{m}\omega_{\mathrm{grav}}^{\mathrm{CS}},\nonumber \\
\int_{\mathrm{TN}_{m}^{\mathrm{sph}}}e_{3}^{(m)}\wedge\omega^{\mathrm{CS}}\left(X\right) & = & \frac{2}{m}A_{L}\wedge F_{L},\\
\int_{\mathrm{TN}_{m}^{\mathrm{sph}}}\chi_{3}\wedge\omega^{\mathrm{CS}}\left(\hat{\omega}_{\mathrm{sph}}\right) & = & \frac{1+4r'+10r'^2+8r'^3+2r'^4}{m\left(1+r'\right)^4} \times 16\pi^{2}\chi_{3},\nonumber 
\end{eqnarray}
where we introduced $r'\equiv m_{\infty}r/m.$ Using that 
\begin{equation}
\chi_{3}=\frac{1}{16\pi^{2}}A_{L}\wedge F_{L}+\frac{1}{8\pi^{2}}\omega^{\mathrm{CS}}\left(A_{R}\right),
\end{equation}
we find that (\ref{eq:integral appendix}) becomes
\begin{eqnarray}
\int_{\mathrm{TN}_{m}^{\mathrm{sph}}}\hat{G}^{\alpha}\wedge\hat{\omega}_{\mathrm{grav}}^{\mathrm{CS}} & = & Q^{\alpha}\left(2\pi\right)^{2}\Biggl[\omega_{\mathrm{grav}}^{\mathrm{CS}}-\frac{1+4r'+2r'^2}{\left(1+r'\right)^4}A_{L}\wedge F_{L}\nonumber \\
 &  & +2\frac{1+4r'+10r'^{2}+8r'^{3}+2r'^{4}}{\left(1+r'\right)^{4}}\omega^{\mathrm{CS}}(A_{R})\Biggl].
\end{eqnarray}
This leads to the expression (\ref{higher derivative result}).

\section{6D to 5D one-loop corrections \label{sec:6D-to-5D}}

In this appendix we use the index theorems to reproduce the results
of \cite{Bonetti:2013ela} in which they calculated the one-loop corrections
that one gets when integrating out massive chiral particles after
the reduction from six to five dimensions on a circle. The gauge field that is relevant
in this case is the $\mathfrak{u}(1)$ Kaluza-Klein vector $A^{0}.$
We will calculate the contributions from spin-$\frac{1}{2}$ fermions,
spin-$\frac{3}{2}$ fermions and for anti-symmetric tensors separately.

\paragraph*{Spin-$\boldsymbol{\frac{1}{2}}$ fermions.}

We consider a massive spin-$\frac{1}{2}$ fermion coupled to the gauge
field $A^{0}$ and to an external gravitational field denoted by the
vielbein $e$. The parity anomaly resulting from this particle can
be canceled by a term 
\begin{equation}
-i\pi \,\mathrm{sign}(M)\int_{\mathcal{M}_{5}}Q(A^{0},\omega),\label{counterterm spin 1/2 6 to 5}
\end{equation}
where 
\[
\upd Q(A^{0},\omega)=\hat{A}(\mathcal{M}_{5})\wedge\mathrm{ch}(F^{0})|.
\]
Using that
\begin{equation}
\hat{A}(\mathcal{M}_{5})\wedge\mathrm{ch}(F^{0})|=\frac{1}{\left(2\pi\right)^{3}}\left(\frac{i^{3}}{6}F^{0}\wedge F^{0}\wedge F^{0}+\frac{i}{48}F^{0}\wedge \tr \cR \wedge \cR\right),
\end{equation}
we find that the counterterms should be given by

\begin{equation}
\frac{\mathrm{sign}(M)}{8\pi^{2}}\left(\frac{-1}{6}A^{0}\wedge F^{0}\wedge F^{0}+\frac{1}{48}A^{0}\wedge \tr \cR\wedge \cR \right).
\end{equation}
Comparing conventions in \cite{Bonetti:2013ela} and \cite{AlvarezGaume:1984nf}
we find that we need $A^{0}\rightarrow qiA^{0}$ in the counterterm
above. We also need to do a Wick rotation to obtain a Lorentzian action
which gives another factor $i.$ We thus find the counterterms
\begin{equation}
\frac{\mathrm{sign}(M)}{8\pi^{2}}\left(-\frac{q^{3}}{6}A^{0}\wedge F^{0}\wedge F^{0}-\frac{q}{48}A^{0}\wedge \tr \cR\wedge \cR \right).\label{counter terms spin 1/2 6d to 5d}
\end{equation}

\paragraph*{Spin-$\boldsymbol{\frac{3}{2}}$ fermions.}

For spin-$\frac{3}{2}$ fermions the anomaly is canceled by a term
of the form (\ref{counterterm spin 1/2 6 to 5}) where 
\begin{eqnarray}
\upd Q(A,\omega) & = & \hat{A}(\mathcal{M}_5)\wedge\left(\mathrm{tr \,e^{i\cR/(2\pi)}-1}\right)\wedge\mathrm{ch}(F^{0})|\nonumber \\
 & = & \frac{1}{\left(2\pi\right)^{3}}\left(\frac{5i^{3}}{6}F^{0}\wedge F^{0}\wedge F^{0}-\frac{19i}{48}F^{0}\wedge \tr \cR\wedge \cR \right).
\end{eqnarray}
The counterterms to the Lorentzian action are thus given by 
\begin{equation}
\frac{\mathrm{sign}(M)}{8\pi^{2}}\left(\frac{-5q^{3}}{6}A^{0}\wedge F^{0}\wedge F^{0}+\frac{19q}{48}A^{0}\wedge \tr \cR \wedge \cR \right).\label{counter terms spin 3/2 6d to 5d}
\end{equation}

\paragraph*{Anti-symmetric tensors.}

In this case we were unable to find an index theorem in the literature,
but following the arguments in the main text we postulate that the
relevant index is given by
\begin{equation}
\mathrm{ind}\, iD_{A}=\frac{1}{2}\int L(M)\wedge\mathrm{ch}(2F^{0})|.\label{index self-dual tensor-1}
\end{equation}
Since the tensors are bosons, the counterterm is now given by \cite{AlvarezGaume:1984dr}
\begin{equation}
i\pi\mathrm{sign}(M)\int_{\mathcal{M}_{5}}Q(A^{0},\omega)\label{counterterm self-dual-1}
\end{equation}
where
\begin{eqnarray}
\upd Q(A^{0},\omega) & = & \frac{1}{2}L(M)\wedge\mathrm{ch}(2F^{0})|\nonumber \\
 & = & \frac{1}{8\pi^{3}}\left(\frac{2i^{3}}{3}F^{0}\wedge F^{0}\wedge F^{0}-\frac{i}{6}F^{0}\wedge \tr \cR \wedge \cR \right).
\end{eqnarray}
This implies that the counterterms to the Lorentzian action are
\begin{equation}
\frac{\mathrm{sign}(M)}{8\pi^{2}}\left(\frac{2q^{3}}{3}A^{0}\wedge F^{0}\wedge F^{0}-\frac{q}{6}A^{0}\wedge \tr \cR \wedge \cR \right).\label{counter terms tensors 6d to 5d}
\end{equation}

The terms (\ref{counter terms spin 1/2 6d to 5d}), (\ref{counter terms spin 3/2 6d to 5d})
and (\ref{counter terms tensors 6d to 5d}) are precisely the one-loop
contributions of table 2.2 in \cite{Bonetti:2013ela}. 

\section{Summation of 6D to 3D one-loop corrections \label{sec:Summation-of-6D}}

To perform the sum of the one-loop corrections over the Kaluza-Klein
spectrum, we use the regularization procedure described in the main
text. We need the sums

\begin{eqnarray}
\sum_{j_{L}=\frac{1}{2}mk}^{\infty}1 & = & \frac{1}{2}-\frac{1}{2}km, \quad \quad \quad
\sum_{j_{L}=\frac{1}{2}mk}^{\infty}j_{L}  =  \frac{1}{24}\left(-2+6km-3k^{2}m^{2}\right),\nn \\
\sum_{j_{L}=\frac{1}{2}mk}^{\infty}j_{L}^{2}  &=&  \frac{1}{24}\left(-2km+3k^{2}m^{2}-k^{3}m^{3}\right),\label{eq:sums over j_l}
\end{eqnarray}
where the sum is over integers (half integers) when $\frac{1}{2}mk$
is integer (half integer). The sums used in this section can then
be calculated using (\ref{zeta function regularization}):

\begin{eqnarray}
\sum_{k=1}^{\infty}\sum_{j_{L}=\frac{1}{2}mk}^{\infty}1 & = & -\frac{1}{4}+\frac{1}{24}m, \quad \quad \quad \quad \quad \; \,
\sum_{k=1}^{\infty}\sum_{j_{L}=\frac{1}{2}mk}^{\infty}j_{L} =  \frac{1}{24}-\frac{1}{48}m,\nonumber \\
\sum_{k=1}^{\infty}\sum_{j_{L}=\frac{1}{2}mk}^{\infty}j_{L}^{2} & = & -\frac{m^{3}}{24\cdot120}+\frac{m}{144}, \quad \quad \quad
\sum_{k=1}^{\infty}\sum_{j_{L}=\frac{1}{2}mk}^{\infty}k^{2} = -\frac{1}{2}\frac{m}{120}.
\end{eqnarray}

We calculate the corrections $\Delta k_{L}^{\mathrm{loop}},$
$\Delta k_{R}^{\mathrm{loop}}$ to the levels and the correction $\Delta\left(c_{L}-c_{R}\right)^{\mathrm{loop}}$
separately. Using the projection condition $j_{L}^{3}=\frac{1}{2}mk$,
we calculate each time first the contribution of the $k=0$ representations
and after that the contribution of $k\neq0.$ Since the structure of
representations for small values of $j_{L},$ $j_{R}$ becomes more
complicated, we first calculate the corrections for $m\geq3$
and do the cases $m=1,2$ separately. 

\subsection{Corrections for $m \geq 3$}

We list the $k=0$ representations where all the sums are over integers
\begin{itemize}
\item Spin-$\frac{3}{2}$:
\[
2\bigoplus_{j_{L}=1}^{\infty}\big(j_{L},j_{L}\pm\tfrac{1}{2}\big)^{\mp}.
\]
\item Spin-$\frac{1}{2}$:
\begin{eqnarray*}
2\bigoplus_{j_{L}=2}^{\infty}\big(j_{L},j_{L}\pm\tfrac{3}{2}\big)^{\mp} & \oplus & 2\bigoplus_{j_{L}=0}^{1}\big(j_{L},j_{L}+\tfrac{3}{2}\big)^{-}\oplus2\bigoplus_{j_{L}=1}^{\infty}\big(j_{L},j_{L}\pm\tfrac{1}{2}\big)^{\pm}\\
\oplus2\big(0,\tfrac{1}{2}\big)^{+} & \oplus & 2\big(n_{T}+n_{H}\big)\bigoplus_{j_{L}=1}^{\infty}\big(j_{L},j_{L}\pm\tfrac{1}{2}\big)^{\pm}\oplus2\left(n_{T}+n_{H}\right)\big(0,\tfrac{1}{2}\big)^{+}.
\end{eqnarray*}
\item vectors:
\[
\bigoplus_{j_{L}=1}^{\infty}\left(j_{L},j_{L}\pm1\right)^{\mp}\oplus\left(0,1\right)^{-}\oplus n_{T}\bigoplus_{j_{L}=1}^{\infty}\left(j_{L},j_{L}\pm1\right)^{\pm}\oplus n_{T}\left(0,1\right)^{+}.
\]
\end{itemize}
Note that the $\left(0,1\right)^{-}\oplus n_{T}\left(0,1\right)^{+}$ vector representations are mapped to itself when applying the reality condition. Hence their contribution comes with an extra factor $\frac{1}{2}$.

When $k>0$ the projection condition gives that $j_{L}^{3}=\frac{1}{2}mk\geq\frac{m}{2}$
which means that when $m\geq3$ we only need the representations (again
the sums go with integer steps)
\begin{itemize}
\item Spin-$\frac{3}{2}$:
\[
2\bigoplus_{j_{L}=\frac{1}{2}mk}^{\infty}\big(j_{L},j_{L}\pm\tfrac{1}{2}\big)^{\mp}.
\]
\item Spin-$\frac{1}{2}$:
\[
2\bigoplus_{j_{L}=\frac{1}{2}mk}^{\infty}\big(j_{L},j_{L}\pm\tfrac{3}{2}\big)^{\mp}\oplus2\bigoplus_{j_{L}=\frac{1}{2}mk}^{\infty}\big(j_{L},j_{L}\pm\tfrac{1}{2}\big)^{\pm}\oplus2\left(n_{T}+n_{H}\right)\bigoplus_{j_{L}=\frac{1}{2}mk}^{\infty}\big(j_{L},j_{L}\pm\tfrac{1}{2}\big)^{\pm}.
\]
\item vectors:
\[
\bigoplus_{j_{L}=\frac{1}{2}mk}^{\infty}\left(j_{L},j_{L}\pm1\right)^{\mp}\oplus n_{T}\bigoplus_{j_{L}=\frac{1}{2}mk}^{\infty}\left(j_{L},j_{L}\pm1\right)^{\pm}.
\]
\end{itemize}

\paragraph*{Correction to left level.}

In this case we do not have a contribution of the $k=0$ modes. We
thus only have to calculate the $k>0$ modes and we will do this separately
for the various types of fields contributing to the left level. For the spin-$\frac{3}{2}$ fermions we
get
\begin{eqnarray}
\alpha_{L}^{\mathrm{(3/2)}} & = & 2\sum_{k=1}^{\infty}\sum_{j_{L}=\frac{1}{2}mk}^{\infty}\frac{3}{8\pi}\big(\tfrac{1}{2}mk\big)^{2}\big[2\big(j_{L}-\tfrac{1}{2}\big)+1-2\big(j_{L}+\tfrac{1}{2}\big)-1\big]\nonumber \\
 & = & -\frac{3m^{2}}{8\pi}\sum_{k=1}^{\infty}\sum_{j=\frac{1}{2}mk}^{\infty}k^{2}=\frac{1}{8\pi} \frac{m^3}{80}.\label{spin 3/2 contri to left level}
\end{eqnarray}
In the same way the spin-$\frac{1}{2}$ fermions give
\begin{eqnarray}
\alpha_{L}^{\mathrm{(1/2)}} & = & \frac{1}{8\pi}\sum_{k=1}^{\infty}\sum_{j_{L}=\frac{1}{2}mk}^{\infty} \Big[-12\big(\tfrac{1}{2}mk\big)^{2}+4\big(\tfrac{1}{2}mk\big)^{2} +4\left(n_{T}+n_{H}\right)\big(\tfrac{1}{2}mk\big)^{2}\Big]\nonumber \\
 & = & -\frac{m^{2}}{8\pi}\left(2-n_{T}-n_{H}\right)\sum_{k=1}^{\infty}\sum_{j_{L}=\frac{1}{2}mk}^{\infty}k^{2} 
 =  \frac{1}{8\pi}\frac{m^{3}}{240}\left(2-n_{T}-n_{H}\right).\label{spin 1/2 contri to left level}
\end{eqnarray}
The vectors contribute with
\begin{equation}
\alpha_{L}^{\mathrm{(vect)}}=4\left(1-n_{T}\right)\sum_{k=1}^{\infty}\sum_{j_{L}=\frac{1}{2}mk}^{\infty}\frac{1}{4\pi}\big(\tfrac{1}{2}mk\big)^{2}=-\frac{1}{8\pi}\frac{m^{3}}{120}\left(1-n_{T}\right).\label{tensor contri to left level}
\end{equation}
Adding the contributions (\ref{spin 3/2 contri to left level}), (\ref{spin 1/2 contri to left level})
and (\ref{tensor contri to left level}), we find
\begin{eqnarray}
\Delta k_{L}^{\mathrm{loop}} & = & 8\pi\cdot\left(\alpha_{L}^{\mathrm{(3/2)}}+\alpha_{L}^{\mathrm{(1/2)}}+\alpha_{L}^{\mathrm{(vect)}}\right)\nonumber \\
 & = & -\frac{m^{3}}{8}c_{1}(B)^2,
\end{eqnarray}
where we used the identities (\ref{anomaly and calabi-yau identity}).

\paragraph*{Correction to right level.}

We get the contribution of the $k=0$ modes by summing over the representations
listed above. We first do this for the $2\left(j_{L},j_{L}\pm\frac{1}{2}\right){}^{\mp}$
representations for the spin-$\frac{3}{2}$ fermions, the $2\left(j_{L},j_{L}\pm\frac{3}{2}\right){}^{\mp}$
representations for the spin-$\frac{1}{2}$ fermions and the $\left(1-n_{T}\right)\left(j_{L},j_{L}\pm1\right){}^{\mp}$
representations for the vectors, which are in this order given by
\begin{align}
-\frac{1}{2\pi}\sum_{j_{L}=1}^{\infty}\Big[-\big(j_{L}+\tfrac{1}{2}\big)\big(j_{L}+\tfrac{3}{2}\big)\big(2j_{L}+2\big)+\big(j_{L}-\tfrac{1}{2}\big)\big(j_{L}+\tfrac{1}{2}\big)\big(2j_{L}\big)\Big] & 
=  -\frac{5}{8\pi},\nonumber \\
-\frac{1}{6\pi}\sum_{j_{L}=2}^{\infty}\Big[-\big(j_{L}+\tfrac{3}{2}\big)\big(j_{L}+\tfrac{5}{2}\big)\big(2j_{L}+4\big)+\big(j_{L}-\tfrac{3}{2}\big)\big(j_{L}-\tfrac{1}{2}\big)\big(2j_{L}-2\big)\Big] & = -\frac{83}{8\pi},\nonumber \\
\frac{1}{6\pi}\left(1-n_{T}\right)\sum_{j_{L}=1}^{\infty}\Big[-\left(j_{L}+1\right)\left(j_{L}+2\right)\left(2j_{L}+3\right)+\left(j_{L}-1\right)j_{L}\left(2j_{L}-1\right)\Big] & = \frac{2}{3\pi}\left(1-n_{T}\right).
\end{align}
For the spin-$\frac{1}{2}$ fields we then also need to sum over the other infinite towers of states, namely
the $2\left(j_{L},j_{L}\pm\frac{1}{2}\right)^{\pm}$ and $2\left(n_{T}+n_{H}\right)\left(j_{L},j_{L}\pm\frac{3}{2}\right){}^{\mp}$
representations. These can be determined
by inserting the right relative factors in the first of the sums above. We also add the contributions from the isolated representations, which are not part of an infinite tower in the spectrum.
These are in the case of spin-$\fr{1}{2}$ fields the $2\left(0,\frac{3}{2}\right)^{-}\oplus2\left(1,\frac{5}{2}\right)^{-}\oplus2\left(n_{T}+n_{H}+1\right)\left(0,\frac{1}{2}\right)^{+}$
representations. Their contribution is given by
\begin{equation}
\frac{5}{12\pi}-\left(n_{T}+n_{H}\right)\frac{11}{24\pi}\,.
\end{equation}
Lastly, we need to sum over the isolated $\left(0,1\right)^{-}\oplus n_{T}\left(0,1\right)^{+}$
representations for the vectors. Since they are mapped to itself when applying the reality condition, we have to add an extra factor $\frac{1}{2}$. This results in
\begin{equation}
-\frac{1-n_{T}}{2\pi\,}.
\end{equation}
Summing all the different contributions gives
\begin{equation}
\alpha_{R}^{k=0}=\frac{1}{8\pi}\left(3-\frac{1}{3}n_{H}-\frac{5}{3}n_{T}\right).\label{k=00003D0 contribution to right level}
\end{equation}

We calculate the $k\neq0$ contributions in the same way as for the
left level.\textbf{ }The spin-$\frac{3}{2}$ fermions contribute
\begin{align}
\alpha_{R}^{\mathrm{(3/2)}}&=-\frac{1}{2\pi}\sum_{k=1}^{\infty}\sum_{j_{L}=\frac{1}{2}mk}^{\infty}\Big[-\big(j_{L}+\tfrac{1}{2}\big)\big(j_{L}+\tfrac{3}{2}\big)\big(2j_{L}+2\big)+\big(j_{L}-\tfrac{1}{2}\big)\big(j_{L}+\tfrac{1}{2}\big)2j_{L}\Big]  \nonumber \\
&=\frac{1}{2\pi}\sum_{k=1}^{\infty}\sum_{j_{L}=\frac{1}{2}mk}^{\infty}\Big(\frac{3}{2}+6j_{L}+6j_{L}^{2}\Big) 
=\frac{1}{4\pi}\Big(-\frac{1}{4}-\frac{m}{24}-\frac{m^{3}}{240}\,\Big).\label{spin-3/2 right level}
\end{align}

We calculate the contribution of the spin-$\frac{1}{2}$ fermions
by first summing over the\textbf{ }$\left(j_{L},j_{L}\pm\frac{3}{2}\right)^{\mp}$
representations
\begin{align}
\ax_R^{(1/2)}=&-\frac{1}{6\pi}\sum_{k=1}^{\infty}\sum_{j_{L}=\frac{1}{2}mk}^{\infty}\Big[-\big(j_{L}+\tfrac{3}{2}\big)\big(j_{L}+\tfrac{5}{2}\big)\big(2j_{L}+4\big)+\big(j_{L}-\tfrac{3}{2}\big)\big(j_{L}-\tfrac{1}{2}\big)\left(2j_{L}-2\right)\Big]\nonumber\\
&-\fr{1}{6\pi} \big(1+n_T+n_H \big)\sum_{k=1}^{\infty} \!\sum_{j_{L}=\frac{1}{2}mk}^{\infty}\!\!\!\Big[\big(j_{L}+\tfrac{1}{2}\big)\big(j_{L}+\tfrac{3}{2}\big)\big(2j_{L}+2\big)-\big(j_{L}-\tfrac{1}{2}\big)\big(j_{L}+\tfrac{1}{2}\big)2 j_L\Big]\nonumber\\
=&\fr{1}{4\pi} \Big(-\frac{9}{4}+\frac{7 m}{24}-\frac{m^3}{240} \Big)-\frac{1}{4\pi}\left(1+n_{T}+n_{H}\right)\Big(-\frac{1}{12}-\frac{m}{72}-\frac{m^3}{720} \Big)\, .\lab{spin 1/2 contribution to level}
\end{align}

Lastly, the vectors give
\begin{align}
\alpha_{R}^{\mathrm{(vect)}}&=\frac{1}{6\pi}\left(1-n_{T}\right)\sum_{k=1}^{\infty}\sum_{j_{L}=\frac{1}{2}mk}^{\infty}\Big[-\left(j_{L}+1\right)\left(j_{L}+2\right)\left(2j_{L}+3\right)+\left(j_{L}-1\right)j_{L}\left(2j_{L}-1\right)\Big]\nonumber \\
&=-\frac{1}{6\pi}\left(1-n_{T}\right)\sum_{k=1}^{\infty}\sum_{j_{L}=\frac{1}{2}mk}^{\infty}\left(6+12j_{L}+12j_{L}^{2}\, \right) 
=\frac{1}{4\pi}\left(n_{T}-1\right)\Big(-\fr{2}{3}+\frac{m}{18}-\frac{m^{3}}{360}\,\Big).\label{tensor contribution to right level}
\end{align}
Adding the contributions (\ref{k=00003D0 contribution to right level}),
(\ref{spin-3/2 right level}), (\ref{spin 1/2 contribution to level})
and (\ref{tensor contribution to right level}), we get the following
correction to the right level
\begin{eqnarray}
\Delta k_{R}^{\mathrm{loop}} & = & 4\pi\left[\alpha_{R}^{{k=0}}+\alpha_{R}^{{(3/2)}}+\alpha_{R}^{{(1/2)}}+\alpha_{R}^{\mathrm{(vect)}}\right]\nonumber \\
 & = & \frac{m^{3}}{24}c_{1}(B)^{2}+\frac{m}{3}c_{1}(B)^{2}+m-c_{1}(B)^{2}-14.
\end{eqnarray}

\paragraph*{Correction to $\boldsymbol{c_{L}-c_{R}}$.}

We do these calculations in the same way as before. We first calculate
the contribution of the $k=0$ states. Summing over these states in the spin-$\frac{1}{2}$, spin-$\frac{3}{2}$ and vector spectrum results in
\begin{align}
\ax_{\rm grav}^{k=0}&=\frac{1}{192\pi}\Big[-12\sum_{j_{L}=2}^{\infty}1+4\left(1+n_{T}+n_{H}\right)\sum_{j_{L}=1}^{\infty}1-2\cdot4-2\cdot6+2\cdot2+2\left(n_{H}+n_{T}\right)\cdot2\Big]\nonumber\\
&+4\sum_{j_{L}=1}^{\infty}\frac{7}{64\pi}-4\left(1-n_{T}\right)\sum_{j_{L}=1}^{\infty}\frac{1}{48\pi}-\frac{1}{2}\cdot\frac{1}{48\pi}\left(1-n_{T}\right)\cdot3\nonumber\\
&=\frac{1}{96\pi}(n_{H}+n_{T})-\frac{7}{32\pi}+\frac{1}{96\pi}(1-n_{T})=-\frac{1}{96\pi}\left(20-n_{H}\right).\label{k=00003D0 to gravitational}
\end{align}

We now calculate the contribution of the $k\neq0$ representations.
Spin-$\frac{3}{2}$ fermions give the following correction
\begin{equation}
\alpha_{\mathrm{grav}}^{\mathrm{(3/2)}}=\frac{7}{16\pi}\sum_{k=1}^{\infty}\sum_{j_{L}=\frac{1}{2}mk}^{\infty}1=\frac{7}{16\pi}\Big(-\frac{1}{4}+\frac{1}{24}m\,\Big)=\frac{1}{96\pi}\Big(-\fr{21}{2}+ \fr{7}{4}m\Big)\,.\label{spin-3/2 to gravitational}
\end{equation}
From spin-$\frac{1}{2}$ fermions we obtain
\begin{align}
\alpha_{\mathrm{grav}}^{\mathrm{(1/2)}}&=-\frac{1}{4\cdot48\pi}\sum_{k=1}^{\infty}\sum_{j_{L}=\frac{1}{2}mk}^{\infty}\Big[2 \cdot6-2\cdot2-2\left(n_{T}+n_{H}\right)\cdot 2\Big] \nn\\
&=-\frac{1}{96\pi}\big(4-2\left(n_{T}+n_{H}\right)\big)\sum_{k=1}^{\infty}\sum_{j=\frac{1}{2}mk}^{\infty}1 = -\frac{1}{96\pi}\left(4-2\left(n_{T}+n_{H}\right)\right)\left(-\frac{1}{4}+\frac{1}{24}m\right).\label{spin-1/2 to gravitational}
\end{align}
Finally, for the vectors we find
\begin{equation}
\alpha_{\mathrm{grav}}^{(\mathrm{vect)}}=-\frac{1}{48\pi}\left(1-n_{T}\right)\sum_{k=1}^{\infty}\sum_{j_{L}=\frac{1}{2}mk}^{\infty}4=-\frac{1}{96\pi}\left(1-n_{T}\right)\left(-2+\frac{m}{3}\right).\label{tensor to gravitational}
\end{equation}
Adding the results (\ref{k=00003D0 to gravitational}), (\ref{spin-3/2 to gravitational}),
(\ref{spin-1/2 to gravitational}) and (\ref{tensor to gravitational})
yields the following correction to $c_{L}-c_{R}$
\begin{eqnarray}
\Delta\left(c_{L}-c_{R}\right) & = & 96\pi\cdot\left(\alpha_{\mathrm{grav}}^{{k=0}}+\alpha_{\mathrm{grav}}^{{(3/2)}}+\alpha_{\mathrm{grav}}^{{(1/2)}}+\alpha_{\mathrm{grav}}^{\mathrm{(vect)}}\right)\nonumber \\
 & = & 6m+\left(2m+17\right)c_{1}(B)^{2}-44\, .
\end{eqnarray}

\subsection{Corrections for $m=2$}

For $m=2$ the representations we need to take into account for $k=0$
stay the same. For $k>0$ only the spin-$\frac{1}{2}$ contribution
changes. Summing the correct representations in the spin-$\frac{1}{2}$
sector, we again find the one-loop corrections.

\paragraph*{Correction to left level.}

For the spin-$\frac{1}{2}$ fermions we now find
\begin{align}
\alpha_{L}^{\mathrm{(1/2)}} & =  -12\sum_{k=2}^{\infty}\sum_{j_{L}=k}^{\infty}\frac{1}{8\pi}k^{2}-12\sum_{j_{L}=2}^{\infty}\frac{1}{8\pi}-12\frac{1}{8\pi}+4\sum_{k=1}^{\infty}\sum_{j_{L}=k}^{\infty}\frac{1}{8\pi}k^{2}\nonumber\\
   &+4\left(n_{T}+n_{H}\right)\sum_{k=1}^{\infty}\sum_{j_{L}=k}^{\infty}\frac{1}{8\pi}k^{2}
  =  \frac{1}{120\pi}-\frac{n_{T}+n_{H}}{240\pi}\,,
\end{align}
which is exactly the same as (\ref{spin 1/2 contri to left level})
for $m=2.$ 

\paragraph*{Correction to right level.}

The change in contribution of the spin-$\frac{1}{2}$ fermions is
caused by the\textbf{ }$\left(j_{L},j_{L}\pm\frac{3}{2}\right)^{\mp}\oplus2\left(1,\frac{5}{2}\right)^{-}$
representations. Its contribution is given by
\begin{eqnarray}
-\frac{1}{6\pi}\sum_{k=2}^{\infty}\sum_{j_{L}=k}^{\infty}\Big[-\big(j_{L}+\tfrac{3}{2}\big)\big(j_{L}+\tfrac{5}{2}\big)\big(2j_{L}+4\big)+\big(j_{L}-\tfrac{3}{2}\big)\big(j_{L}-\tfrac{1}{2}\big)\big(2j_{L}-2\big)\Big]\nonumber \\
-\frac{1}{6\pi}\sum_{j_{L}=2}^{\infty}\Big[-\big(j_{L}+\tfrac{3}{2}\big)\big(j_{L}+\tfrac{5}{2}\big)\big(2j_{L}+4\big)+\big(j_{L}-\tfrac{3}{2}\big)\big(j_{L}-\tfrac{1}{2}\big)\big(2j_{L}-2\big)\Big]+\frac{35}{4\pi},
\end{eqnarray}
which when combined with the other spin-$\frac{1}{2}$ representations
(for which the summation goes the same as in the $m\geq3$ case) gives
\begin{equation}
\alpha_{{R}}^{{(1/2)}}=-\frac{71}{180\pi}+\frac{11\left(n_{H}+n_{T}\right)}{360\pi}.
\end{equation}
This is again the same as the contribution (\ref{spin 1/2 contribution to level})
for $m=2$. 

\paragraph*{Correction to $\boldsymbol{c_{L}-c_{R}}$.}

This time, we find the following contribution for the spin-$\frac{1}{2}$
fermions:
\begin{align}
\alpha_{\mathrm{grav}}^{\mathrm{(1/2)}}&=-\frac{1}{4\cdot48\pi}\bigg[2\sum_{k=2}^{\infty}\sum_{j_{L}=k}^{\infty}6+2\sum_{j_{L}=2}^{\infty}6+12-2\sum_{k=1}^{\infty}\sum_{j_{L}=k}^{\infty}2-2\left(n_{T}+n_{H}\right)\sum_{k=1}^{\infty}\sum_{j_{L}=k}^{\infty}2\bigg]  \nonumber \\
&=\frac{1}{144\pi}-\frac{n_{T}+n_{H}}{288\pi},
\end{align}
which is the same as (\ref{spin-1/2 to gravitational}) for $m=2$.

\subsection{Corrections for $m=1$}

The change in summations is again only for $k>0,$ but in this case
it is both in the spin-$\frac{1}{2}$ sector and in the vector sector. 

\paragraph*{Correction to left level.}

For the spin-$\frac{1}{2}$ fermions we find
\begin{align}
\alpha_{L}^{\mathrm{(1/2)}} & =  -12\sum_{k=3}^{\infty}\sum_{j_{L}=\frac{1}{2}k}^{\infty}\frac{1}{8\pi}\big(\tfrac{1}{2}k\big)^{2}-12\sum_{j_{L}=2}^{\infty}\frac{1}{8\pi}-12\sum_{j_{L}=\frac{3}{2}}^{\infty}\frac{1}{8\pi}\left(\frac{1}{2}\right)^{2}-\frac{3}{2\pi}-10\frac{1}{8\pi}\left(\frac{1}{2}\right)^{2}\nonumber \\
&+4\sum_{k=2}^{\infty}\sum_{j_{L}=\frac{1}{2}k}^{\infty}\frac{1}{8\pi}\left(\frac{1}{2}k\right)^{2}+4\sum_{j_{L}=3/2}^{\infty}\frac{1}{8\pi}\left(\frac{1}{2}\right)^{2}+6\frac{1}{8\pi}\left(\frac{1}{2}\right)^{2}\nonumber \\
&+4\left(n_{T}+n_{H}\right)\sum_{k=1}^{\infty}\sum_{j_{L}=\frac{1}{2}k}^{\infty}\frac{1}{8\pi}\left(\frac{1}{2}k\right)^{2} =  \frac{121}{960\pi}-\frac{n_{H}+n_{T}}{1920\pi},
\end{align}
which is not the same as (\ref{spin 1/2 contri to left level}) for
$m=1.$
The vector contribution changes to

\begin{eqnarray}
\alpha_{L}^{\mathrm{(vect)}} & = & 4\left(1-n_{T}\right)\sum_{k=2}^{\infty}\sum_{j_{L}=\frac{1}{2}k}^{\infty}\frac{1}{4\pi}\left(\frac{1}{2}k\right)^{2}+4\left(1-n_{T}\right)\sum_{j_{L}=\frac{3}{2}}^{\infty}\frac{1}{4\pi}\cdot\frac{1}{4}+4\left(1-n_{T}\right)\frac{1}{4\pi}\cdot\frac{1}{4}\nonumber \\
 & = & -\frac{1-n_{T}}{960\pi},
\end{eqnarray}
which is exactly the same as (\ref{tensor contri to left level})
for $m=1.$ 

\paragraph*{Correction to right level.}

We first calculate the contribution of the spin-$\frac{1}{2}$ fermions.
For the $2\bigoplus_{j_{L}=\frac{3}{2}}^{\infty}\left(j_{L},j_{L}\pm\frac{3}{2}\right)^{\mp}$
representations we find
\begin{align}
&-\frac{1}{6\pi}\bigg[\sum_{k=3}^{\infty}\sum_{j_{L}=k/2}^{\infty}+\sum_{j_{L}=2}^{\infty}+\sum_{j_{L}=3/2}^{\infty}\bigg]\Big[-\big(j_{L}+\tfrac{3}{2}\big)\big(j_{L}+\tfrac{5}{2}\big)\big(2j_{L}+4\big)
+\big(j_{L}-\tfrac{3}{2}\big)\big(j_{L}-\tfrac{1}{2}\big)\big(2j_{L}-2\big)\Big] \nonumber \\
&=-\frac{4557}{320\pi}.
\end{align}
The $2\bigoplus_{j_{L}=1}^{\infty}\left(j_{L},j_{L}\pm\frac{1}{2}\right)^{\pm}$
representations give
\begin{align}
&-\frac{1}{6\pi}\bigg[\sum_{k=2}^{\infty}\sum_{j_{L}=k/2}^{\infty}+\sum_{j_{L}=3/2}^{\infty}\bigg]\Big[-\big(j_{L}+\tfrac{1}{2}\big)\big(j_{L}+\tfrac{3}{2}\big)\big(2j_{L}+2\big)
+\big(j_{L}-\tfrac{1}{2}\big)\big(j_{L}+\tfrac{1}{2}\big)\big(2j_{L}\big)\Big]\nonumber \\& =
\frac{2951}{2880\pi}.
\end{align}
For the $2\bigoplus_{j_{L}=0}^{1}\left(j_{L},j_{L}+\frac{3}{2}\right)^{-}\oplus2\left(\frac{1}{2},1\right)^{+}$
representations we find a contribution
\begin{equation}
-\frac{1}{6\pi}\left(-\frac{5}{2}\cdot\frac{7}{2}\cdot6-2\cdot3\cdot5+1\cdot2\cdot3\right)=\frac{51}{4\pi}.
\end{equation}
The contribution of the $2\left(n_{T}+n_{H}\right)\bigoplus_{j_{L}=\frac{1}{2}}^{\infty}\left(j_{L},j_{L}\pm\frac{1}{2}\right)^{\pm}$
representations stays the same. Adding all the contributions we find
\begin{equation}
\alpha_{{R}}^{{(1/2)}}=-\frac{671}{1440\pi}+\left(n_{H}+n_{T}\right)\frac{71}{2880\pi},
\end{equation}
which is the same as (\ref{spin 1/2 contribution to level}) for $m=1.$
The vector contribution changes to
\begin{align}
\alpha_{R}^{\mathrm{(vect)}} & =  \frac{1}{6\pi}\left(1-n_{T}\right)\sum_{k=2}^{\infty}\sum_{j_{L}=\frac{1}{2}k}^{\infty}\left[-\left(j_{L}+1\right)\left(j_{L}+2\right)\left(2j_{L}+3\right)+\left(j_{L}-1\right)j_{L}\left(2j_{L}-1\right)\right]\nonumber \\
 &  +\frac{1}{6\pi}\left(1-n_{T}\right)\sum_{j_{L}=\frac{3}{2}}^{\infty}\left[-\left(j_{L}+1\right)\left(j_{L}+2\right)\left(2j_{L}+3\right)+\left(j_{L}-1\right)j_{L}\left(2j_{L}-1\right)\right]\nonumber \\
 & = \left(1-n_{T}\right)\frac{221}{1440\pi},
\end{align}
which is the same as (\ref{tensor contribution to right level}) for
$m=1$.

\paragraph*{Correction to $\boldsymbol{c_{L}-c_{R}}$.}

Now we find for the spin-$\frac{1}{2}$ fields
\begin{align}
\alpha_{\mathrm{grav}}^{\mathrm{(1/2)}} & =  \frac{1}{4\cdot48\pi}\Biggl[-2\sum_{k=3}^{\infty}\sum_{j_{L}=k/2}^{\infty}6-2\sum_{j_{L}=2}^{\infty}6-2\sum_{j_{L}=3/2}^{\infty}6-12-10\nonumber \\
 & \qquad \qquad ~~\,    +2\sum_{k=2}^{\infty}\sum_{j_{L}=k/2}^{\infty}2+2\sum_{j_{L}=3/2}^{\infty}2+6-2\left(n_{T}+n_{H}\right)\sum_{k=1}^{\infty}\sum_{j_{L}=k/2}^{\infty}2\Biggl]\nonumber \\
 & =  \frac{17}{576\pi}-\frac{5}{1152\pi}\left(n_{T}+n_{H}\right).
\end{align}
For the vectors we get
\begin{align}
\alpha_{\mathrm{grav}}^{(\mathrm{vect)}} & =  -\frac{1}{12\pi}\left(1-n_{T}\right)\bigg[\sum_{k=2}^{\infty}\sum_{j_{L}=\frac{1}{2}k}^{\infty}1+\sum_{j_{L}=3/2}^{\infty}1+1\bigg] =  \frac{5}{288\pi}-\frac{5}{288\pi}n_{T},
\end{align}
which is the same as (\ref{tensor to gravitational}) for $m=1$.

\bibliographystyle{utcaps}
\bibliography{references}

\end{document}